\documentclass[11pt]{article}
\usepackage[letterpaper,margin=1in]{geometry}
\usepackage{graphicx}
\usepackage{longtable,booktabs,array,calc}
\usepackage{amsmath,amssymb}
\usepackage{enumitem}
\usepackage{xurl}
\usepackage[hidelinks]{hyperref}
\usepackage{microtype}
\usepackage{caption}
\usepackage{float}
\usepackage{etoolbox}
\AtBeginEnvironment{longtable}{\scriptsize}
\graphicspath{{figures/}}
\title{A Solid-Solid Phase Transition Reverses Carrier Selectivity in the Molecular Crystal Electrolyte (Adpn)$_2$LiI}
\author{Shujit Chandra Paul$^{1}$, Giselle de Araujo$^{2}$, Phillip Stallworth$^{2}$, Steven Greenbaum$^{2}$,\\
William A. Goddard III$^{3}$, Prabhat Prakash$^{3,*}$, Stephanie Wunder$^{1,*}$, Michael Zdilla$^{1,*}$\\[0.75em]
\small $^{1}$Department of Chemistry, Temple University, Philadelphia, PA 19122, USA\\
\small $^{2}$Department of Physics, Hunter College, CUNY, New York, New York 10065, USA\\
\small $^{3}$Materials and Process Simulation Center (MSC), California Institute of Technology,\\
\small Pasadena, California 91125, USA}
\date{}
\begin{document}
\let\savedaddcontentsline\addcontentsline
\renewcommand{\addcontentsline}[3]{}
\maketitle
\begin{abstract}

Understanding and controlling temperature-dependent phase transitions in
molecular solid electrolytes is essential for achieving high lithium-ion
conductivity. Here we show that thermally induced structural
reorganization in the adiponitrile--lithium iodide
(Adpn)\textsubscript{2}LiI molecular crystal drastically enhances Li\textsuperscript{+}
transport and can be kinetically trapped using 5\% isostructural
(Adpn)\textsubscript{2}LiPF\textsubscript{6}. The low-temperature (LT),
lower conductivity phase (5 x 10\textsuperscript{-5} S/cm at 25
\textsuperscript{0}C) adopts an orthorhombic structure with channels
with relatively large (6.1 Å)
Li\textsuperscript{+}···Li\textsuperscript{+} distances and where
Li\textsuperscript{+} ions are blocked by anions. Upon heating, blocked
channels in alternate layers reorganize via translation in opposite
directions, transforming the LT to a tetragonal high-temperature (HT)
phase with shorter (5.2 Å) Li\textsuperscript{+}
···Li\textsuperscript{+} distances in unblocked channels and an order of
magnitude higher conductivity (3.5 x 10\textsuperscript{-4} S/cm at 25
\textsuperscript{0}C), and a rise in the Li\textsuperscript{+}
self-diffusion coefficient from 5.15 \ensuremath{\times} 10\textsuperscript{-11} to 8.1 \ensuremath{\times}
10\textsuperscript{-11} m\textsuperscript{2} s\textsuperscript{-1} near
the transition temperature. The Li\textsuperscript{+} transference number increases from
0.10 to 0.65 in the unblocked channels, confirming a shift from
anion-dominated to cation-dominated transport.

\end{abstract}

\vfill
\begin{center}
\small
\href{mailto:pprakash@caltech.edu}{\texttt{pprakash@caltech.edu}};
\href{mailto:slwunder@temple.edu}{\texttt{slwunder@temple.edu}};
\href{mailto:mzdilla@temple.edu}{\texttt{mzdilla@temple.edu}}
\end{center}
\newpage

\section{Introduction}

The development of solid-state electrolytes for next-generation
batteries is essential for advancing the safety, thermal stability, and
energy density of lithium-ion batteries (LIBs)\textsuperscript{1,2}, but
widespread adoption has been hindered by limited ionic conductivity and
poor interface compatibility\textsuperscript{3}. Ionic transport in
solid electrolytes is inherently structure-dependent, with parameters
such as coordination environment, lattice dynamics, and defect pathways
playing critical roles\textsuperscript{4--9}. Phase changes can have
significant impact on the overall conductivity of solid electrolytes,
with superionic behavior occurring in Li\textsubscript{7}Zn\textsubscript{0}.\textsubscript{5}SiS\textsubscript{6} as the result of a
temperature-induced phase transition from monoclinic to tetragonal
symmetry,\textsuperscript{10} and in LiBH\textsubscript{4} from a
low-temperature (LT) orthorhombic\textsuperscript{11} to a high
temperature (HT) hexagonal phase at 390 K, which raises the conductivity
by three orders of magnitude.\textsuperscript{12} In the HT superionic
phase there are channels for unimpeded Li\textsuperscript{+} ion
migration along two crystallographic axes, and the HT phase can be
stabilized at room temperature by the addition of lithium
halides.\textsuperscript{13} One or two orders of magnitude higher RT
ionic conductivity (10\textsuperscript{-4} S/cm to
10\textsuperscript{-5} S/cm) occur for Li halide
stabilized\textsuperscript{12}
LiI-LiBH\textsubscript{4}\textsuperscript{13} compared with
LiBH\textsubscript{4}.\textsuperscript{14}

Molecular crystal-based solid electrolytes have emerged as promising
candidates for Li\textsuperscript{+}~ion battery (LIB) applications due
to their inherent thermal stability, ease of synthesis, and competitive
ionic conductivity.\textsuperscript{15--17} Temperature-dependent phase
transitions can induce subtle changes in molecular packing or
crystallographic symmetry, which in turn can significantly affect ionic
mobility. Despite this potential, the exploitation of thermally driven
solid--solid phase transitions to enhance ionic conductivity remains
unexplored in this new class of electrolyte materials.

Recently, we reported a molecular crystal-based solid electrolyte
(Adpn)\textsubscript{2}LiI designed for lithium--iodide battery
systems\textsuperscript{7} that undergoes a solid--solid phase
transition near 76\,°C, followed by a sharp melting point at
approximately 130\,°C. In its low-temperature form, the crystal adopts
an orthorhombic structure, where lithium ions are coordinated with
nitrile groups that precludes the presence of well-defined
Li\textsuperscript{+}~ion-conducting channels. Our previous studies
demonstrated that at an adiponitrile-to-LiI molar ratio of 2:1, the
system primarily supports I\textsuperscript{-} ion conduction. However,
increasing the ratio to 3:1 introduces excess Adpn at the grain
boundaries, enabling selective Li\textsuperscript{+}~ion transport
across crystal interfaces.

Irrespective of the Adpn:LiI molar ratio, upon heating above the phase
transition temperature (\textasciitilde76\,°C), the crystals transform
into a tetragonal phase that features more ordered and extended
ion-conducting pathways. This structural rearrangement raises a
fundamental question: does the high-temperature phase, with its more
ordered geometry, enhance lithium-ion mobility through the bulk of the
crystal? To address this, a detailed investigation of the crystal
structure, phase behavior, and temperature-dependent conductivity was
conducted.

The objective of this study is to examine the atomic-level
rearrangements that occur during the phase transition of the Adpn:LiI
co-crystal and to determine how these structural changes influence ionic
conductivity, particularly lithium-ion transport. By combining
crystallographic analysis with impedance and NMR spectroscopy, we aim to
provide insights into the relationship between molecular packing, phase
transitions, and ion conduction in soft-solid electrolytes. Our
investigation clearly demonstrates that the solid--solid phase
transition facilitates the formation of more accessible ion transport
channels for lithium ions, leading to a sharp increase in ionic
conductivity around 70--80\,°C and a concurrent reduction in activation
energy barriers. Doping with 5\% LiPF\textsubscript{6}, whose
(Adpn)\textsubscript{2}LiPF\textsubscript{6} crystal is isostructural with (Adpn)\textsubscript{2}LiI, kinetically
stabilizes the solid-solid phase transition to room temperature.

\section{Results and Discussion}

\subsection{Structural Transformation of the (Adpn)\textsubscript{2}LiI molecular crystal from the Orthorhombic to Tetragonal Phase}

The low-temperature (LT) phase of the (Adpn)\textsubscript{2}LiI
molecular crystal crystallizes in an orthorhombic unit cell with the
Pnna space group (\textbf{Figure 1a})\textbf{,} with lattice parameters
a=20.4001~Å,~b=12.3504~Å,~c=12.3248~Å, and all 90° interaxial angles.
The asymmetric unit contains a Li\textsuperscript{+} ion coordinated by two terminal
nitrile groups from two distinct adiponitrile molecules, while the
remaining two coordination sites are completed through the 2-fold
rotation operation, resulting in a tetrahedral coordination environment
(\textbf{Figure S1a}). Upon heating, a solid--solid phase transition
occurs near 75°C, during which the structure transforms from the
orthorhombic to a tetragonal unit cell adopting the P4/n space group
(\textbf{Figure 1b}). The corresponding unit cell parameters change to
\emph{a'} = \emph{b'} = 12.619~Å and \emph{c'} = 5.2192~Å, while the
angles remain 90°. In this case, the asymmetric unit contains only one
nitrile group coordinated with the Li\textsuperscript{+} ion, while the remaining three
coordination sites are generated through 4-fold rotoinversion and 2-fold
rotation symmetry operations, completing the tetrahedral coordination
geometry (\textbf{Figure S1b}).

\includegraphics[width=\textwidth,height=0.82\textheight,keepaspectratio]{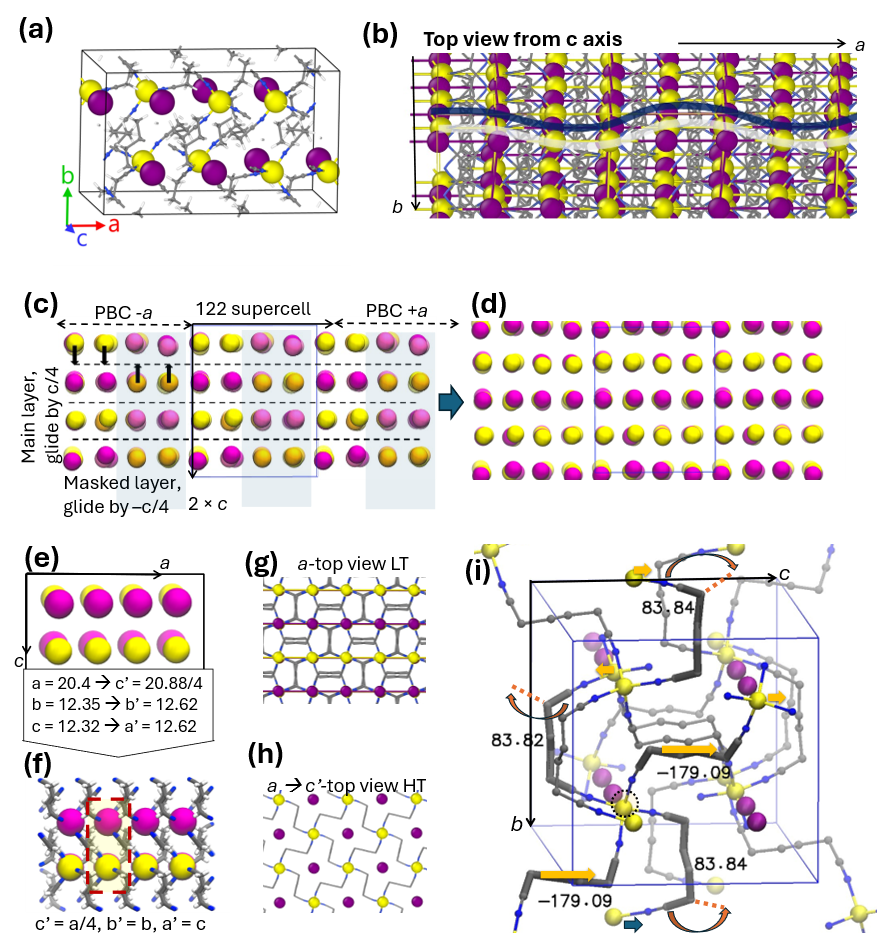}

\textbf{Figure 1. (a)} A c-axis top view of unit cell of
Adpn\textsubscript{2}LiI LT-phase structure; \textbf{(b)} Expanded view
showing distorted channels of Li\textsuperscript{+~}and
I\textsuperscript{-} ions (connecting respective diagonal ions) in LT
phase which are blocked due to the presence of counterions after each
two repetitions; \textbf{(c)} Expanded view of Li\textsuperscript{+}~and
I\textsuperscript{-}~occupancy sites for a model 1 x 2 x 2 supercell
constructed with periodic boundary images shown in +a and -a directions,
proposed glide motion to transform LT to HT is: +c/4 glide for each two
Li\textsuperscript{+} ion layers (main layer), and -c/4 glide for each
alternate -- masked layers, glide motion is shown with small arrows in
+c, -c directions, with dotted line (in this 2D representation)
suggesting the plane of new occupancy sites post transformation;
\textbf{(d)} Post glide-transformed supercell structure; \textbf{(e)}
Extracted LT-unit-cell equivalent slice of transformed structure shown
in (d); \textbf{(f)} Extracted HT-unit cell equivalent (bordered with
red dotted line) obtained as c' = a/4, b' = b, and a' = c conversion of
axis notation; \textbf{(g)} Low-conductive Li\textsuperscript{+} and
I\textsuperscript{-} channels that are present in LT phase in its b
crystallographic direction, which are shown removed in \textbf{(h),} as
better, high-conductive, channels form in c' (= a) crystallographic
direction of HT phase; \textbf{(i)} All absolute structural changes that
convert LT to HT phase, straight orange arrows show glide movement for
Li\textsuperscript{+} ions relative to a central reference
Li\textsuperscript{+} ion (marked with black dotted circle) for ions to
rearrange from LT to HT; Curved arrows show the direction of dihedral
rotation of Adpn required for two Adpn molecules to change their
dihedral from \textasciitilde85 to \textasciitilde{} 180 degrees, with
new expected C-CN bonds shown in orange dotted lines. (yellow)/light
orange Li, (purple)/Magenta I, (gray) C, (blue) N, white H. a, b, c
are axis notations for LT phase, a', b', c' are axis notations for HT
phase.

A closer examination of the LT phase reveals that Li\textsuperscript{+}
(and I\textsuperscript{-}) ions that exist as alternate successive
duet-neighbors
(\ldots Li\textsuperscript{+}\ldots Li\textsuperscript{+}\ldots I\textsuperscript{-}\ldots I\textsuperscript{-}\ldots)\textsubscript{n}
in the \emph{a}-crystallographic direction of the orthorhombic cell
\textbf{(Figure 1a)} have Li\textsuperscript{+}\ldots{}
Li\textsuperscript{+},
I\textsuperscript{-}\ldots I\textsuperscript{-}~and
Li\textsuperscript{+}\ldots I\textsuperscript{-}~distances of 5.1 ± 0.1
Å. These Li\textsuperscript{+}\ldots Li\textsuperscript{+} and
I\textsuperscript{-}\ldots I\textsuperscript{-}~are ``blocked''
channels, since two counterions block the direction of linear movement
of the ions parallel to the \emph{a-}crystallographic direction.
However, it can be seen from the \emph{c-axis} top view \textbf{(Figure
1b)}, that distorted channels could form for both Li\textsuperscript{+}
and I\textsuperscript{-} ions, but only with a long-diagonal
intermediate jump of 8 Å with the respective neighbor ions at a relative
location of ±a/4 + c/2 or ±a/4 - c/2 (absolute coordinate units). A
requirement to hop through such long-distances limits ion-conduction in
the \emph{a-}crystallographic direction, and thus hopping conduction is
seen to occur in the \emph{b-}crystallographic direction with a
Li\textsuperscript{+}\ldots Li\textsuperscript{+} distance of 6.2 Å.
However, thermal fluctuations can glide these diagonal layers such that
Li\textsuperscript{+}~and I\textsuperscript{+}~ions form alternate
channels (separated by \emph{c}/2) in the \emph{a-}crystallographic
direction. \textbf{Figure 1c} shows a model structure simulated in a
molecular dynamics (MD) framework, which forces
Li\textsuperscript{+}~and I\textsuperscript{-}~ions in the main-layer
(nominal convention) to glide by \emph{c}/4, and each alternate
distorted layer (shown as a masked layer) to glide by \emph{-c}/4/ in
the c-crystallographic direction, such that, the alternate successive
neighbors now form separate channels of Li\textsuperscript{+} and
I\textsuperscript{-} in the \emph{a-}crystallographic direction
\textbf{(Figure 1d)}, with
Li\textsuperscript{+}\ldots Li\textsuperscript{+} =
I\textsuperscript{-}\ldots I\textsuperscript{-} = 5.1 ± 0.1 Å distances.

An ordered slice of the transformed 122 supercell was constructed to
mimic a transformed LT unit-cell \textbf{(Figure 1e),} that has \emph{a
=} 20.4 Å, \emph{b =} 12.35 Å, \emph{c =} 12.32 Å. The transformed
structure is not a reducible unit-cell structure, as it shows four
periodic repetitions in the \emph{a}-crystallographic direction. Further
reduction of the structure in \textbf{Figure 1e} to ¼ x 1 x 1 cell
\textbf{(Figure 1f)}, creates a unit cell with \emph{a =} 5.1 Å, \emph{b
=} 12.35 Å, \emph{c =} 12.32 Å, which resembles very closely the HT
unit-cell dimensions of \emph{a' = b' =}12.619~Å and \emph{c' =}
5.2192~Å (using \emph{a'} = \emph{c}, \emph{b}\textsuperscript{'} = b,
and \emph{c' = a)} which are within error bars, 2 -- 3 \%, in each
dimension the cell.

\textbf{Figure 1g} and \textbf{h} show the disappearance of the old
Li\textsuperscript{+}~and I\textsuperscript{-}~channels in the LT phase,
as it transforms to the HT phase with more ordered channels in the
\emph{c}-crystallographic direction. The proposed model of the LT to HT
conversion is also consistent with the change in Adpn end-to-end
distances observed in the single crystal data of the LT and HT phases.
There are two types of Adpn molecules in the LT phase:

\begin{enumerate}
\def\labelenumi{\roman{enumi})}
\item
  Two of four Adpn that are coordinated to each Li\textsuperscript{+}
  ion have an end-to-end (N-N) distance of 5.8 Å, with a dihedral
  C(1)N-C(2)-C(5)-C(6)N \textasciitilde{} 80 degrees,
\item
  The other two Adpn of the four coordinated to each
  Li\textsuperscript{+} ion have end-to-end distance of 6.9 Å, with a
  dihedral C(1)N-C(2)-C(5)-C(6)N \textasciitilde{} 180 degrees.
\end{enumerate}

In the HT phase, all four Adpn molecules have a constant 6.9 Å
end-to-end distance with a dihedral angle close to 180 degrees. Thus,
the LT to HT conversion also changes type-i to type-ii Adpn molecules.
\textbf{Figure 1i} shows that uniformity in Adpn conformers is a
phenomenological change, associated with c-glide motion responsible for
the transformation, as relative c/4 and -c/4 glides of two diagonal (0,
0, 0) and (a/4, 0, c/2) Li\textsuperscript{+} ions would result in an
absolute change of c/2, which can be obtained by the rotation of type-i
Adpn shown with the yellow colored curved arrow. These observations
indicate that during the solid--solid transition, both Li\textsuperscript{+} and I\textsuperscript{-} ions
undergo relocation within the unit cell, forming an ordered channel in
the \emph{a} (of LT, = \emph{c} of HT) -crystallographic direction with
the disappearance of the old channels that existed in the
\emph{b}-crystallographic direction of the LT phase. This glide pathway,
constructed here by hand, is modeled by the solid-state NEB calculations
described later. Additionally, because of the clear change in crystal
arrangement between the LT and HT phases, the predicted powder
diffraction patterns differ significantly for the two structures. The
experimental PXRD data for both phases closely match their respective
simulated patterns, confirming that the bulk crystal retains its
structural identity until the phase transition occurs \textbf{(Figure
2a)}.

\includegraphics[width=\textwidth,height=0.82\textheight,keepaspectratio]{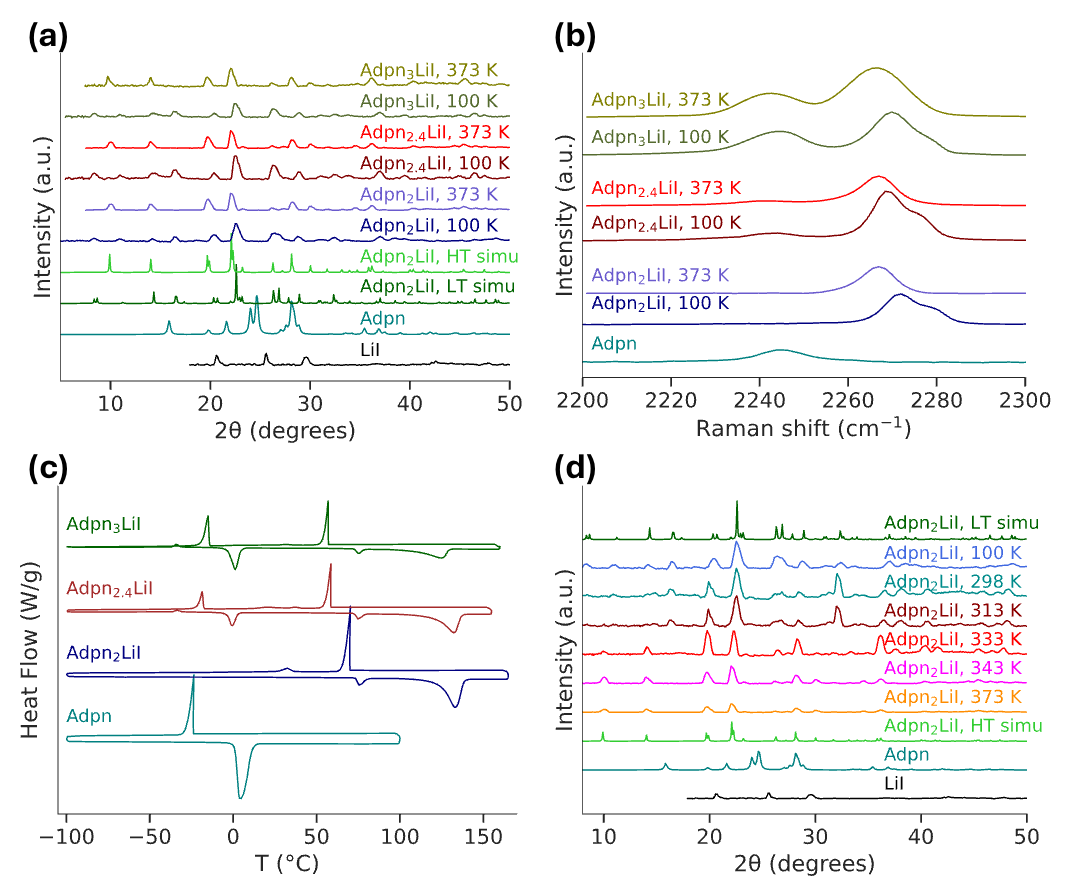}

\textbf{Figure 2.} Comparison of \textbf{(a)} PXRD data and \textbf{(b)}
Raman data for both HT and LT phases with Adpn:LiI stoichiometries of
2:1, 2.4:1 and 3:1. In PXRD data, simu = simulated PXRD patterns
obtained from single crystal data. \textbf{(c)} DSC data of the various
composition showing the LT-HT phase transition \textbf{(d)} Temperature
dependent PXRD data of the 2:1 Adpn:LiI sample.

Raman spectroscopy data supports the change in environment for
Li\textsuperscript{+} ions during the LT to HT transformation. When
Li\textsuperscript{+} ions are coordinated to Adpn, the characteristic
C\ensuremath{\equiv}N stretching peak for neat Adpn at \textasciitilde{} 2245 cm\textsuperscript{-}\textsuperscript{1}
undergoes a blue shift to a single sharp peak at \textasciitilde{} 2275
cm\textsuperscript{-1} for (Adpn)\textsubscript{2}LiPF\textsubscript{6}
molecular crystals.\textsuperscript{8,15} For LT-phase of
Adpn\textsubscript{2}LiI, a main C\ensuremath{\equiv}N stretching peak was observed at
2273 cm\textsuperscript{-}\textsuperscript{1} accompanied by a shoulder peak at approximately 2280 cm\textsuperscript{-}\textsuperscript{1}
(\textbf{Figure 2b}), which is due to the existence of two conformers of
Adpn molecules in two different chemical
environments.\textsuperscript{7} Temperature-dependent Raman
measurements collected at 85\,°C show a single, sharper peak at 2266
cm\textsuperscript{-}\textsuperscript{1}, indicating a structural reorganization into a more uniform
coordination environment for Li\textsuperscript{+} ions in the HT
phase.. However, this Li\textsuperscript{+}--nitrile (--CN) coordination band is red
shifted in the HT compared to the LT phase, implying improved geometric
regularity in the HT tetragonal phase.

\subsection{Thermal Properties}

The polymorphic transition of the Adpn:LiI molecular crystal occurs
without passing through the liquid or gas phase, and differential
scanning calorimetry (DSC) is a reliable technique for detecting this
kind of transition. DSC analysis reveals a distinct exothermic
solid--solid phase transition, T\textsubscript{ss} \textasciitilde75\,°C
for the (Adpn)\textsubscript{2}LiI molecular crystal. For the
non-stoichiometric crystals, 2.4:1 and 3:1, the additional adiponitrile
accumulates at the grain boundaries\textsuperscript{8} and leads to a
decrease in T\textsubscript{ss} by approximately 1\,°C (\textbf{Figure
2c}), but the enthalpy (\ensuremath{\Delta}H\textsubscript{ss}) of the transition remains
constant (4.75\,kJ/mol) for all molar ratios. Upon cooling, a reversible
endothermic peak to the orthorhombic phase was observed near 35°C. This
observation confirms that the solid-solid phase transition is repeatable
upon thermal cycling. A similar solid--solid phase transition
(T\textsubscript{ss} = 115\,°C) has been reported for LiBH\textsubscript{4} with an
associated enthalpy change of approximately
4.15\,kJ/mol\textsuperscript{18}

The retention of the HT phase is further supported by
temperature-dependent PXRD data (\textbf{Figure 2d}), where the HT phase
persists down to approximately 40\,°C, while a transition to the LT
phase is observed when the data is collected at 20\,°C. To determine
whether the phase transition is time-dependent or purely
temperature-driven, we performed DSC measurements at different cooling
rates for the 2:1 Adpn:LiI molecular crystal (Figure S2a). The results
clearly indicate that the transition temperature from the tetragonal to
the orthorhombic phase is cooling-rate dependent, shifting from
approximately 35 °C to around 55 °C when the cooling rate is decreased
to 1 °C/min. The DSC result differs from the temperature dependent PXRD
data because the PXRD data was taken more quickly compared to the DSC
experiments. The melt temperature, T\textsubscript{m}, of the tetragonal
phase is 132\,°C, and gradually decreases with increasing amount of
excess adiponitrile in the grain boundaries, which depresses the melting
point.

\subsection{Electrochemical properties}

\includegraphics[width=\textwidth,height=0.82\textheight,keepaspectratio]{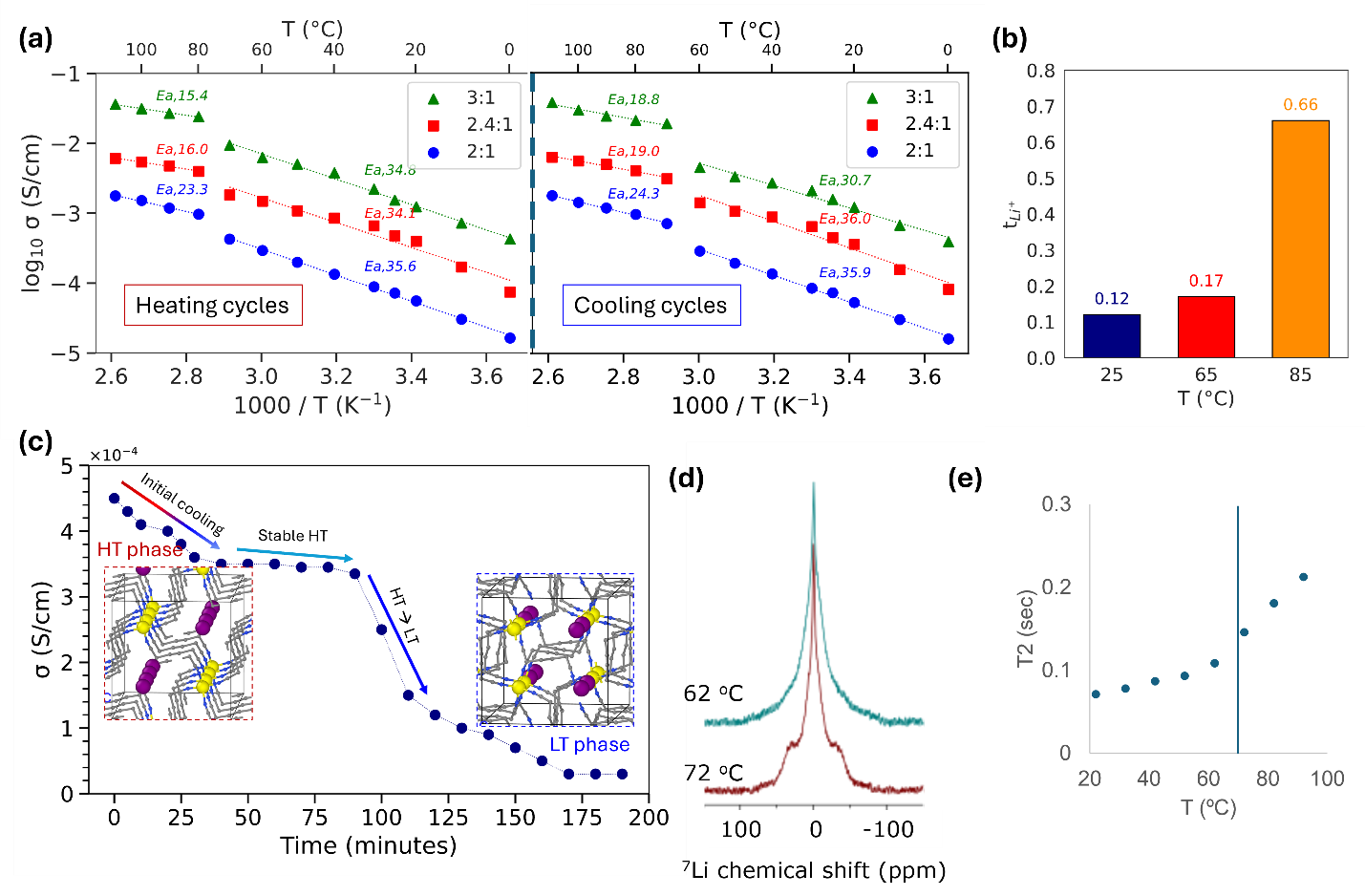}

\textbf{Figure 3.} Temperature dependent conductivity data for
(Adpn)\textsubscript{2}LiI in the \textbf{(a)} Heating cycle, and
\textbf{(b)} Cooling cycle; Arrhenius activation energy barriers for two
segments of temperature dependence are provided in
kJ-mol\textsuperscript{-1}; (\textbf{c}) Time dependent conductivity
data of (Adpn)\textsubscript{2}LiI doped with 5 mol \% isomorphic
(Adpn)\textsubscript{2}LiPF\textsubscript{6} after heating to 125
\textsuperscript{0}C and quenching to 25 \textsuperscript{0}C; (yellow)
Li, (purple) I, (gray) C, (blue) N; (\textbf{d}) Temperature dependent
transference number (t\textsubscript{Li}\textsuperscript{+}) data for
(Adpn)\textsubscript{2}LiI; \textbf{(d)} Comparison of static
\textsuperscript{7}Li spectra at 62 and 72 °C of a sample of
(Adpn)\textsubscript{3}LiI, highlighting the change in quadrupolar
powder pattern across the solid--solid transition; \textbf{(e)}
T\textsubscript{2} times of the mobile species as a function of
temperature. The vertical line indicates the onset of the solid-solid
phase transition (\textasciitilde70 °C).

Crystallographic analysis reveals that the molecular crystal undergoes a
significant structural rearrangement during the solid-solid phase
transition, which is expected to have a pronounced impact on its overall
ionic conductivity. Temperature-dependent ionic conductivity
(\textbf{Figure 3a}) shows a continuous increase in ionic conductivity
with temperatures up to 70\,°C for all compositions. However, at T
\textgreater{} 70 \textsuperscript{0}C a distinct discontinuous jump is
observed, with conductivity increasing sharply from 4.2 \ensuremath{\times} 10\textsuperscript{-}\textsuperscript{3} S/cm at
70\,°C to 9.6 \ensuremath{\times} 10\textsuperscript{-}\textsuperscript{3} S/cm at 80\,°C. This sudden enhancement is
attributed to the structural reorganization that results in a shorter
Li--Li distance, consistent with our previous
findings\textsuperscript{5,19}, thereby lowering the activation energy
(E\textsubscript{a}) barrier from E\textsubscript{a} = 35 kJ/mol for T
\textless{} 70 \textsuperscript{0}C to E\textsubscript{a} =24 kJ/mol for
T \textgreater{} 80 °C. Notably, in the 2.4:1 and 3:1 Adpn:LiI
compositions, the reduction in activation energy is even more pronounced
(about 16 kJ-mol\textsuperscript{-1} for 2.4:1 molecular crystal), due
to the presence of excess adiponitrile at the grain boundaries, which
facilitates selective lithium-ion migration and further reduces the
energy barrier for ion transport. When cooled from the melt the samples
supercool and the discontinuous jump in conductivity shifts to around
60--70\,°C as the tetragonal HT phase persists to lower temperature
before transitioning to the orthorhombic phase, as confirmed by the
rate-dependent DSC measurements (Figure S2a). Although the increase in
conductivity between the LT and HT phases is smaller than that reported
for LiBH\textsubscript{4},\textsuperscript{12} the conductivity of the
HT phase (at 384K) is the same as that of the HT LiBH\textsubscript{4}
phase (at 390K).\textsuperscript{12}

To verify that the observed rise in conductivity is not a trivial
thermal effect, we attempted to prolong the HT phase on cooling by
lowering the solid--solid transition temperature through anion doping.
Anion doping, by addition of 5\% LiPF\textsubscript{6}, prolonged the HT phase to lower
temperatures during the cooling cycle. Since (Adpn)\textsubscript{2}LiPF\textsubscript{6} and
(Adpn)\textsubscript{2}LiI\textsubscript{6} are isostructural (same space group and unit-cell type), our
working hypothesis was that introducing PF\textsubscript{6}\textsuperscript{-} would retard
ion-reorganization kinetics and shift the HT\ensuremath{\leftrightarrow}LT transition to lower
temperature. Indeed, undoped (Adpn)\textsubscript{2}LiI transforms to the LT phase near
\textasciitilde35 °C, whereas the 5\% LiPF\textsubscript{6}-doped crystal undergoes the
phase transition at --40 °C (\textbf{Figure S2a}), expanding the
isostructural window by an additional \textasciitilde75 °C for
\textasciitilde{} 2 hours. This broader window allowed measurement of
the RT ionic conductivity of the HT phase before it reverted to the LT
phase. A 5\% LiPF\textsubscript{6} doped (Adpn)\textsubscript{2}LiI\textsubscript{6} sample was melted in a coin cell at
approximately 125 °C and quenched to 25 °C at 20 °C/min. The
conductivity (\textbf{Figure 3c}) of the electrolyte was 4.5 \ensuremath{\times}
10\textsuperscript{-}\textsuperscript{4} S cm\textsuperscript{-}\textsuperscript{1} immediately after quenching (first EIS
measurement, 0 min), stabilized to 3.5 \ensuremath{\times} 10\textsuperscript{-}\textsuperscript{4} S cm\textsuperscript{-}\textsuperscript{1} as the system
reached thermal equilibrium and remained constant for about 40--50
minutes. At the phase transition to the LT phase, the conductivity
dropped to 0.5 x 10\textsuperscript{-\/-4} S/cm, almost an order of
magnitude lower in conductivity than the HT phase at the same (RT)
temperature.

The reduction in Li···Li distances should also directly influence the
lithium-ion transference number
(\emph{t}\textsubscript{Li}\textsuperscript{+}).\textsuperscript{5}
Temperature-dependent measurements of
\emph{t}\textsubscript{Li}\textsuperscript{+} (\textbf{Figure 3b and
S3a-c}) indicate that in the LT phase, there is no significant change in
\emph{t}\textsubscript{Li}\textsuperscript{+}, with nearly identical
values observed at room temperature and 65\,°C. However, at 85\,°C,
\emph{t}\textsubscript{Li}\textsuperscript{+} sharply increases from
approximately 0.17 to 0.66, higher than for the previously reported the
succinonitrile:LiPF\textsubscript{6} molecular crystal (Li-Li distance
of 5.73 Å)\textsuperscript{5}, due the smaller Li-Li distance in the HT
phase (5.2 Å). Comparison of the Li/Li cell resistance before
polarization reveals a 23\% decrease at 65\,°C and a more pronounced
75\% reduction at 85\,°C, further supporting the proposed hypothesis of
reduced resistance to Li\textsuperscript{+} ion migration
(\textbf{Figure S3d}).

In our previous study\textsuperscript{7} the 2:1 Adpn:LiI molecular
crystal was shown to behave predominantly as an anion conductor;
however, through grain boundary (GB) engineering, cyclic voltammetry
(CV) data showed transformation into a cation-conducting electrolyte.
This transition was evidenced by a significant increase in the lithium
transference number and a notable change in the redox behavior, from
iodide-dominated chemistry to lithium-based electrochemistry. Since the
present system also involves structural reorganization, we used CV to
test whether similar changes in the redox processes occur. CV analysis
(\textbf{Figure S4}) reveals similar changes in the redox behavior for
the HT phase, with iodide redox reactions prominent in the LT and
lithium plating dominant in the HT phase. These observations strongly
suggest that structural reorganization significantly enhances
lithium-ion mobility within the HT crystal lattice, leading to the
transition from iodide-dominated conduction to lithium-ion conduction.

\subsection{NMR study}

NMR spectroscopic studies of (Adpn)\textsubscript{n}LiI samples support
the conclusion that the major effect of the thermal structural
reorganization is to increase Li\textsuperscript{+} ion mobility within
the grain boundaries.

Static \textsuperscript{7}Li spectra of (Adpn)\textsubscript{3}LiI
acquired between 22 and 92 °C reveal the interplay between structural
ordering and lithium mobility across the solid--solid phase transition
(\textbf{Figure S5a}). At low temperatures (\textless{} 70 °C)
(\textbf{Figure 3d}), the spectra are dominated by a very broad base,
consistent with strong Li--Li dipolar interactions that are not motion
averaged. There is significant quadrupolar broadening that does not
resolve into distinct satellite transitions, suggesting a distribution
of surrounding electric field gradient (EFG) orientations that are
greatly disordered, arising from a distribution of Li environments. This
feature suggests the presence of lithium ions that are largely localized
in rigid environments, giving rise to long relaxation times
(T\textsubscript{1} \ensuremath{\approx} 26 s at 52 °C) and reduced
mobility\textsuperscript{20,21}. Samples with a superstoichiometric
quantity of Adpn also feature a superimposed narrow feature, indicating
that a minority of ions already possess higher mobility. (\textbf{Figure
S5b}) This is consistent with our model in which most of the more
immobile charge carriers are in the grains, but the smaller number of
more mobile (by \textasciitilde{} 2 orders of magnitude) charge carriers
are in the grain boundaries.\textsuperscript{8}

Upon heating to \textasciitilde72 °C, where the molecular crystal
undergoes a solid--solid phase transition, the broad dipolar background
is progressively averaged by lithium hopping, allowing the underlying
quadrupolar interaction to emerge as a well-defined quadrupolar powder
pattern (\textbf{Figure 3d}). In this regime, lithium ions move rapidly
between equivalent sites: the motion is fast enough to average dipolar
couplings but not so fast or isotropic as to completely average the
electric field gradient (EFG). Consequently, quadrupolar interactions
remain partially visible, consistent with behavior reported for fast-ion
conductors.\textsuperscript{20--22} This is also consistent with the
differences between inorganic and molecular crystalline lithium-ion
conductors. For the inorganic materials the higher conductivity occurs
in the grains, with lower conductivity in the resistive grain
boundaries. In the case of molecular crystals, the fast ion conduction
occurs for ions in the fluid nano-confined grain boundaries, with the
slower ion hopping motion of in the channels of the grains. The enhanced
mobility is also manifest in a reduction of the longitudinal relaxation
time (T\textsubscript{1} \ensuremath{\approx} 14 s at 72 °C) of the broad component.

Application of a CPMG pulse train permits isolation of the narrow
Li\textsuperscript{+} signal assigned to the material within the grain
boundaries (\textbf{Figure S5c}), allowing quantitative insight into the
behavior and relative concentration of this critical species at
temperatures above and below the phase transition. These results
indicate that the solid-solid phase transition does not notably increase
the concentration of Li\textsuperscript{+} within the grain boundaries
but does substantially increase the Li\textsuperscript{+} jump rate.
Absolute integration data indicate intensities of the narrow species
increase with temperature but only seem to be limitedly effected by the
solid-solid phase transition (\textbf{Figure S5d}). By contrast,
T\textsubscript{2} values of the mobile species are dramatically
impacted by the solid-solid phase transition (\textbf{Figure 3e}). At
temperatures below the solid-solid phase transition, T\textsubscript{2}
values are nearly temperature independent. Above the transition, the
T\textsubscript{2} values show a strong correlation with temperature.
Similar inflection points have been noted for other lithium conductors
and indicate an increased rate of lithium hopping.\textsuperscript{23}
Increases in T\textsubscript{2} only begins as mobility is sufficient to
overcome the rigid lattice local field. Thus, the solid-solid phase
transition results in a significant increase in the mobility of the
Li\textsuperscript{+} located between the grain boundaries.

\subsection{Mechanisms for Li\textsuperscript{+} (and I\textsuperscript{-}) ion conduction}

The mechanism for Li\textsuperscript{+} and I\textsuperscript{-} ion
conduction in the HT phase is derived using nudged elastic band (NEB)
calculations on vacancy created supercells using plane-wave periodic
density functional theory (PWDFT) calculations. The HT phase shows that
both Li\textsuperscript{+} and I\textsuperscript{-} ions move in
channels in the \emph{c-}crystallographic direction with activation
energy (E\textsubscript{a}) barriers of 0.43 and 0.42 eV, respectively
\textbf{(Figure 4)}. The reaction coordinate (RC) is defined as zero
when the Li\textsuperscript{+} (or I\textsuperscript{-}) is at an
occupancy site (\textbf{Figure 4a,} for Li\textsuperscript{+~}and
\textbf{d} for I\textsuperscript{-}), and as one when
Li\textsuperscript{+} (or I\textsuperscript{-}) reaches the first
vacancy site (\textbf{Figure 4c} for Li\textsuperscript{+} and
\textbf{f} for I\textsuperscript{-}). The minimum energy path (MEP) is
sampled with 9 intermediate images. For a Li\textsuperscript{+} ion, the
most energetic barrier along the reaction coordinate is the initial
breaking of the Li\ldots CN coordination requiring 0.37 eV to reach
\textasciitilde{} RC = 0.3. At this RC, the two CN groups from the
initial occupancy site are left behind, and the two remaining CN groups
carry the Li\textsuperscript{+} ion over the next 40\% of the RC (from
RC = 0.3 to 0.7). At the transition state (TS), where MEP has a peak (at
RC = 0.5, \textbf{Figure 4g}) of 0.43 eV, there is a coordination
reconstruction around the Li\textsuperscript{+} ion, with two CN groups
from the initial (occupancy) site and two CN groups from the final
(vacancy) site \textbf{(Figure 4b)} assist the Li\textsuperscript{+} ion
hopping. This coordination reconstruction is a common mechanisms for
Li\textsuperscript{+} ions to hop in a crystalline interior as
previously seen in several other examples.\textsuperscript{7,15} Adpn
has 11 local minima of structural conformers \textbf{(Figure S6)} from
the rotation of its dihedral backbone alone, providing several possible
values for end-to-end N-N distances (8.1 to 3.8 Å, separated within 3
kcal-mol\textsuperscript{-1}). The ability of the alkyl chain to
\emph{compress} or \emph{extend} contributes to the facilitation or
hindrance for Li\textsuperscript{+} ion hopping.

Activation barriers for a Li\textsuperscript{+} ion in such cases are
affected, structurally, by three factors:

\begin{enumerate}
\def\labelenumi{\roman{enumi})}
\item
  Relative distance between successive Li\textsuperscript{+} ions in the
  channel
\item
  Rotation along the NC-C2-C5-CN dihedral for the Adpn backbone
\item
  Bending in the C-C-CN angle of Adpn
\end{enumerate}

The above factors act in a concerted manner. Factor \emph{i} determines
the extent of stretch along the Adpn backbone required to support the TS
structure, with rotations (factor \emph{ii)} and bends (factor
\emph{iii)} changing to meet the required distance. In the LT phase, the
calculated E\textsubscript{a} barrier from DFT was 0.9
eV.\textsuperscript{7} The significantly lower value of
E\textsubscript{a} (\textsubscript{Li+}) in the HT phase is due to a
(15\%) shorter Li\textsuperscript{+}···Li\textsuperscript{+} distance
(5.2 Å in HT \emph{vs.} 6.18 in LT) that the Adpn backbone needs to
accommodate, ease of dihedral rotation for the Adpn backbone
(\emph{d}\textsubscript{C-C-C-C} is 180 ± 1 in HT vs. 172.5 ±1 in LT),
and less bending stress on the C-C-CN angle
(\emph{a}\textsubscript{C-C-CN} is 117° in T vs. 111° in LT).

\includegraphics[width=\textwidth,height=0.82\textheight,keepaspectratio]{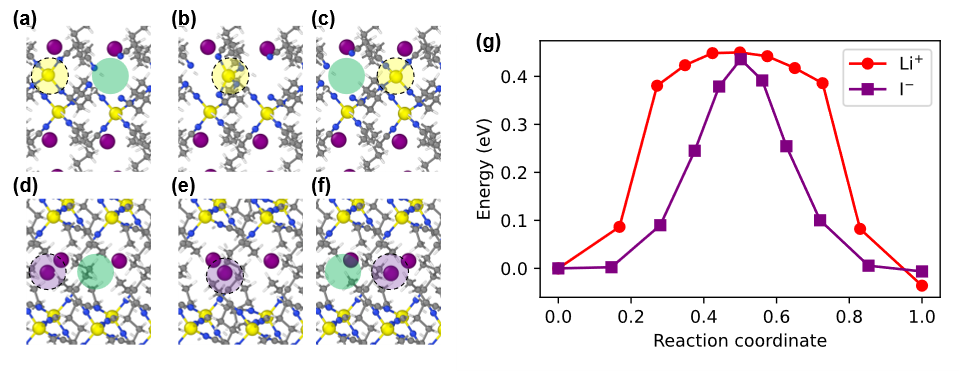}

\textbf{Figure 4.} Minimum energy~path for ion-conduction, derived from
DFT -- NEB calculations, in the high-temperature phase of
(Adpn)\textsubscript{2}LiI molecular crystals for
Li\textsuperscript{+}~ions (\textbf{a}), (\textbf{b}), (\textbf{c}) and
I\textsuperscript{-}~ions (\textbf{d}), (\textbf{e}), (\textbf{f}). The
moving ion is highlighted~with the dotted circles,~vacancy site is
highlighted in green ; (b) and (e) show transition state structures for
both (Li+ and I-, respectively) cases; color coding for atoms:  yellow
Li,, (purple) I, (gray) C, (blue) N; (\textbf{g}) Relative energy
profile for Li+~and I-~conduction in molecular crystals. A video showing
the mechanism is provided in supplementary movie 2.

\subsection{Dynamics of ions during structural transformation}

Molecular dynamics (MD) simulations with classical interatomic
potentials were used to simulate the structural transformation from the
LT to HT phase, starting with a 3x5x5 LT-Adpn\textsubscript{2}LiI
supercell for two temperature ramps of 100 -- 600 K (\textbf{Figure S7})
and 300 -- 400 K (\textbf{Figure S8})\textbf{.} The mass density, which
decreased gradually until 360 K, dropped dramatically in the 360 -- 380
K window. The calculated non-bonded interaction energy (NBIE) profiles
for Adpn and I\textsuperscript{-} around Li\textsuperscript{+} during
this thermal transformation show significant changes between 360 -- 380
K \textbf{(Figure S7b, S8b)}: (i) A Li\textsuperscript{+} ion changes
its coordination from
Li\textsuperscript{+}\ldots{[}N\ensuremath{\equiv}C(Adpn){]}\textsubscript{4}
(\textless NBIE\textgreater{} decreases from \textasciitilde{} -- 62.5
kJ-mol\textsuperscript{-1} to -- 35 kJ-mol\textsuperscript{-1}, per Li-N
non-covalent interaction) to
Li\textsuperscript{+}···I\textsuperscript{-}
(\textless NBIE\textgreater{} increases from -- 12
kJ-mol\textsuperscript{-1} to -- 28 kJ-mol\textsuperscript{-1}); (ii) As
the crystal structure collapses
Li\textsuperscript{+}···I\textsuperscript{-} ion-pairs phase-separate,
with two-to-three Adpn interacting with the
\ldots{[}Li\textsuperscript{+}···I\textsuperscript{-}···Li\textsuperscript{+}{]}\ldots{}
linear wires. The trajectory of Li\textsuperscript{+~}ions
\textbf{(Figure S8c-f)} in the pre-decomposition window (340 -- 360 K)
shows some Li\textsuperscript{+~}ions moving in the regular counterion
occupancy sites, formed via their mutual exchange with
I\textsuperscript{-}~occupancy sites, suggesting a short-lived LT -- HT
transformation, before the structure collapses in the 360 -- 380 K
window (\textbf{Figure S8f, S9}).

Histogram distributions of tracer diffusion coefficients for
Li\textsuperscript{+} (\textbf{Figure S10}) and I\textsuperscript{-}
(\textbf{Figure S11}), calculated from MD trajectories show that the
cage vibrations (10\textsuperscript{-14} \textless{}
\textbf{D\textsubscript{Li}\textsuperscript{+}} \textless{}
10\textsuperscript{-12} in m\textsuperscript{2}-s\textsuperscript{-1})
are much smaller than the diffusion coefficients measured by NMR
linewidth analysis, reaffirming that the NMR diffusion coefficients come
from Li ions in the grain boundaries. As the phase transition begins at
340 K, some carriers also exhibit a
D\textsubscript{Li}\textsuperscript{+} between 10\textsuperscript{-12}
-- 10\textsuperscript{-11} (highlighted with gray color region in
\textbf{Figures S10, S11)}, that corresponds to the interstitial
reorganization of Li\textsuperscript{+}~and I\textsuperscript{-} ions
during structural transformation. The force-field was parameterized for
the LT phase and loses lattice integrity at 360-380 K. These simulations
illustrate the ion exchange that accompanies the transition but are not
evidence for the structure or the stability of the HT phase, which rest
on single-crystal and powder diffraction. A transformation path that
does not depend on this classical force-field is obtained below with a
DFT-validated machine-learned (ML) potential.

\subsection{Grain-boundary transport}

To separate the contributions of the grains and of the grain boundaries,
we simulated a crystal grain in contact with excess Adpn, as in the
non-stoichiometric samples, where the additional Adpn accumulates at the
grain boundaries.\textsuperscript{8} The 3\ensuremath{\times}5\ensuremath{\times}5 LT supercell used above
(600 formula units, \textasciitilde6 nm across) was placed in a periodic
box with 600 to 1200 additional Adpn molecules, giving overall
compositions of (Adpn)\textsubscript{3}LiI and
(Adpn)\textsubscript{4}LiI and boundary layers \textasciitilde1.1 and
\textasciitilde1.7 nm thick between periodic images of the grain. The
excess Adpn was first equilibrated around the fixed grain (5 ns, NVT,
300 K), and the whole system was then simulated without constraints for
10 ns (NpT, 300 K); a second (Adpn)\textsubscript{4}LiI run was heated
from 100 to 300 K over 1 ns before 9 ns at 300 K \textbf{(Supporting
Information)}. In every 1 ns window each Li\textsuperscript{+} ion was
classified as caged, when it kept four nitrile N neighbors and no
I\textsuperscript{-} within 3.3 Å, as in the LT crystal, or as
disordered otherwise, and tracer diffusion coefficients were computed
over the same 1 ns windows as in \textbf{Figures S10 and S11}.

While the grain is intact (the first 2--3 ns), the two populations are
clearly separated (Figure S17, Table S8). Caged ions more than 10 Å
inside the grain show only cage vibrations (median D = 2 - 7 \ensuremath{\times}
10\textsuperscript{-13} m\textsuperscript{2} s\textsuperscript{-1} for
both Li\textsuperscript{+} and I\textsuperscript{-}), as in the bulk
crystal at 300 K \textbf{(Figure S10a)}, whereas ions in the disordered
boundary layer diffuse 40--90 times faster on average (median D =
1.3-3.1 \ensuremath{\times} 10\textsuperscript{-11} m\textsuperscript{2}
s\textsuperscript{-1}), within a factor of 2-6 of the
\textsuperscript{7}Li PFG-NMR values (5.15 \ensuremath{\times} 10\textsuperscript{-11} and
8.1 \ensuremath{\times} 10\textsuperscript{-11} m\textsuperscript{2}
s\textsuperscript{-1}). By the Nernst--Einstein relation, a boundary
population of 5--10\%, as estimated from the static
\textsuperscript{7}Li line shape, then carries 70-90\% of the ionic
conductance, 2-10 times the contribution of the grain interiors
\textbf{(Figure S18e)}. This supports the assignment of the PFG-NMR
diffusion, and of most of the measured conductivity, to the
grain-boundary regions in both phases. In the boundary layer
Li\textsuperscript{+} and I\textsuperscript{-} diffuse at similar rates
(Nernst-Einstein computed t\textsubscript{Li+} \ensuremath{\approx} 0.55). It can be thus
concluded that the classical simulations do not resolve the carrier
selectivity, which is addressed by the NEB barriers above (making those
barriers an important part of total conductivity and carrier selectivity
assessment).

\subsection{Solid-state NEB of the LT \ensuremath{\rightarrow} HT transformation}

The glide model discussed earlier was constructed by hand, and the
classical force-field fails in the transition window, so neither
provides an unbiased transformation path or barrier. We therefore
computed the minimum energy path between the DFT-optimized LT and HT
structures with the variable-cell solid-state NEB
method\textsuperscript{24} and a MACE\textsuperscript{25}
machine-learned interatomic potential (MLIP), fine-tuned from
MACE-MP-0\textsuperscript{26} and validated against DFT data (training
details are in Supporting Information). In the smallest cell that
contains both structures (one LT cell, 8 formula units) all formula
units transform together. The barrier of this concerted path is 0.96 eV
per formula unit from VASP single points on the optimized band (0.98 eV
with the MLIP), with a second maximum of 0.89 eV \textbf{(Figure 5a)}.
The trained MLIP reproduces the DFT energies along the path to 26 meV
per formula unit and the forces to 11--23 meV Å\textsuperscript{-1}.

The derived minimum energy path reproduces the essential features of the
proposed glide mechanism without imposing them \textbf{(Figure 5b)}.
Four of the eight Li\textsuperscript{+}-I\textsuperscript{-} columns
along the LT c axis slide by ±c/2 (6.2 Å), two in each direction, so
that their Li\textsuperscript{+} and I\textsuperscript{-} ions exchange
sites, while the other four columns move by less than 0.6 Å
(Li\textsuperscript{+}) and 1.8 Å (I\textsuperscript{-}). This relative
c/2 displacement of neighboring columns is equivalent, up to a rigid
translation, to the opposed ±c/4 glides of alternate layers in
\textbf{Figure 1c}. Adpn molecules barely translate (their centers move
by \textasciitilde0.2 Å) but rotate: the eight gauche molecules
(C1-C2-C5-C6 dihedral 83°) become trans (180°), while the eight trans
molecules stay trans. This converts the 50:50 gauche:trans mixture of
the LT phase into the all-trans HT phase, as in the single-crystal
structures and thus also merge the two Raman C\ensuremath{\equiv}N components into a
single band \textbf{(Figure 2b)}.

A path in which every formula unit of the crystal transforms at once is
an upper bound, because a macroscopic crystal can always chose a
plausible lower energy path where it can transform by nucleation and
growth. To test sequential conversion, we mapped the band onto a 2\ensuremath{\times}2\ensuremath{\times}2
supercell (2,176 atoms) and let the four rods along c (each 1\ensuremath{\times}1\ensuremath{\times}2 LT
cells, a quarter of the supercell) convert one after another
\textbf{(Figure 5c)}. On the rod-by-rod path the highest image, during
conversion of the second rod, is 0.68 eV/f. u. above LT, 30\% below the
concerted barrier with the same potential (0.98 eV/f.u.), and the states
with one, two and three converted rods lie 0.38, 0.54 and 0.35 eV/f.u.
above LT.

Thus the computed pathway supports the proposed glide mechanism of the
phase transition, and indicate that it proceeds by growth of HT domains
along c rather than by a concerted transformation of the whole crystal.
The energy cost of LT/HT interfaces is consistent with the hysteresis of
the transition (T\textsubscript{ss} \ensuremath{\approx} 75 °C on heating, 35-55 °C on
cooling, \textbf{Figure 2c and S2a}) and with the slow, time-dependent
reversion of the quenched, doped HT phase to LT \textbf{(Figure 3c)}.

\includegraphics[width=\textwidth,height=0.82\textheight,keepaspectratio]{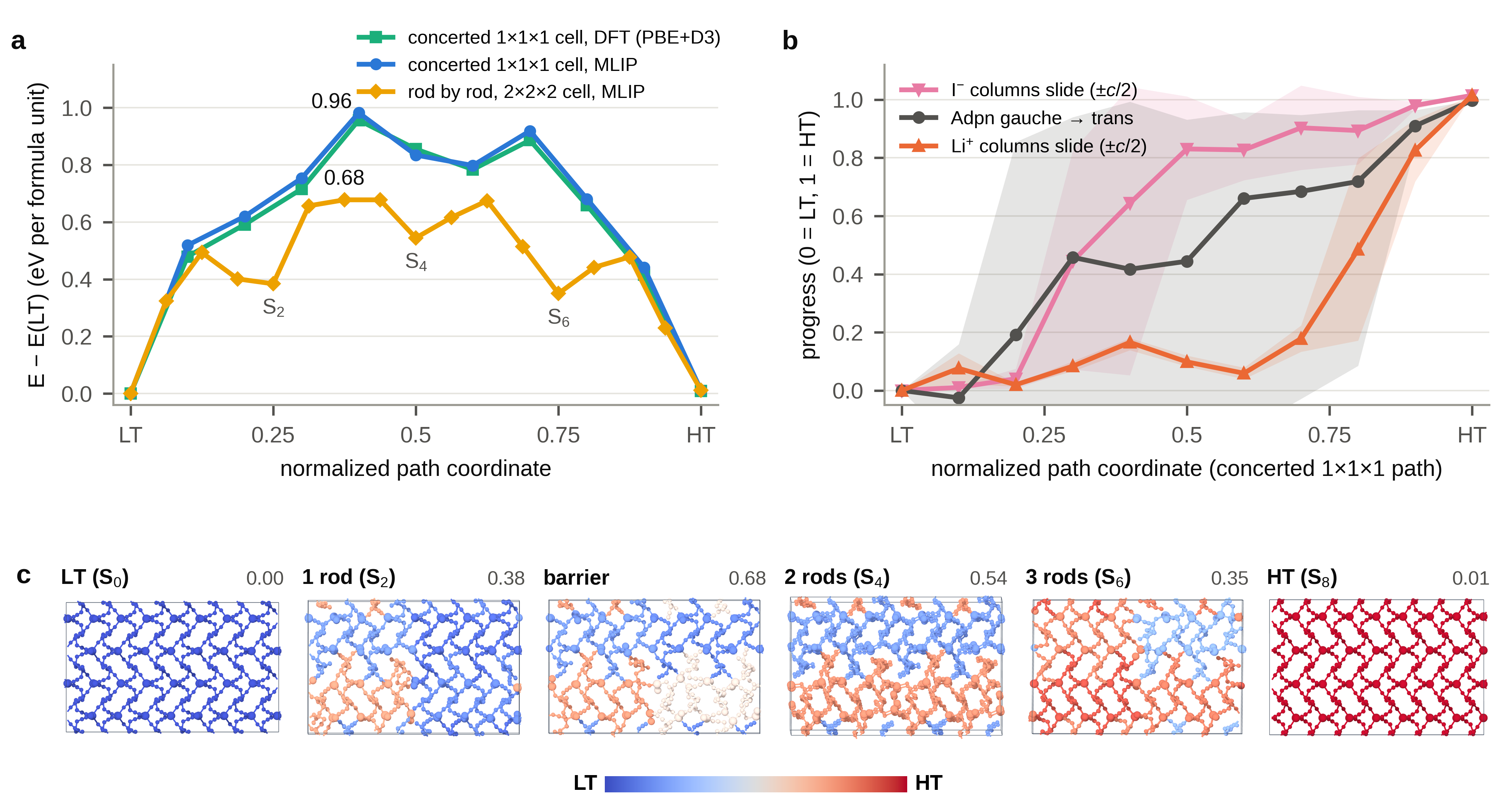}

\textbf{Figure 5.} Solid-state NEB of the LT \ensuremath{\rightarrow} HT transformation of
(Adpn)\textsubscript{2}LiI. (a) Energy along the concerted path in one
LT cell (8 formula units) from the fine-tuned machine-learned
interatomic potential (MLIP, blue) and from DFT (PBE+D3(BJ)) single
points on the same images (green), and along the rod-by-rod path in a
2\ensuremath{\times}2\ensuremath{\times}2 supercell (64 formula units; MLIP, yellow). Energies are relative
to LT, per formula unit of the simulation cell; S\textsubscript{2},
S\textsubscript{4} and S\textsubscript{6} are the states with one, two
and three converted rods. (b) Structural changes along the concerted
path: slide of the four I\textsuperscript{-} and four
Li\textsuperscript{+} columns that move by ±c/2 and rotation of the
eight gauche Adpn molecules to trans ((\ensuremath{\theta} - 83°)/(180° - 83°), \ensuremath{\theta} =
C1-C2-C5-C6 dihedral). Lines are means and shaded bands the range over
the columns or molecules. (c) Views down c of the 2\ensuremath{\times}2\ensuremath{\times}2 supercell
(hydrogens hidden) along the rod-by-rod path, with atoms colored by the
conversion progress of their LT cell (blue, LT; red, HT); numbers of
right top corner are energies in eV per formula unit.

\section{Conclusions}

We demonstrate that the thermally induced solid--solid phase transition
in adiponitrile--lithium iodide (Adpn)\textsubscript{2}LiI molecular
crystals acts as a fundamental "structural switch" that dictates ion
transport efficiency. By transitioning from an orthorhombic
low-temperature phase to a more conductive tetragonal high-temperature
phase at \textasciitilde75°C, the system undergoes a significant lattice
reorganization that creates continuous, shorter distance Li\textsuperscript{+}-conduction
channels. This structural shift results in an order-of-magnitude
increase in ionic conductivity and a dramatic rise in the Li\textsuperscript{+}
transference number from 0.10 to 0.65, effectively shifting the
transport mechanism from anion-dominated to cation-dominated.
Computational and NMR analyses further confirm that the reduction of
Li\textsuperscript{+}--Li\textsuperscript{+} distances to 5.2 Å within these channels lowers the activation
energy barrier for cation hopping and higher cation transference.
Solid-state NEB calculations with a DFT-validated MLIP shows the
proposed glide mechanism as a ±c/2 slide of half of the
Li\textsuperscript{+}-I\textsuperscript{-} columns coupled to the
\emph{gauche} to \emph{trans} rotation of half of the Adpn molecules and
show that rod-by-rod conversion through LT/HT interfaces is favored over
a concerted transformation (\ensuremath{\approx} 0.7 vs \ensuremath{\approx} 1.0 eV/f.u.). The successful
kinetic stabilization of this high-conductivity phase at room
temperature through LiPF\textsubscript{6} doping highlights a viable
pathway for engineering practical, high-performance electrolytes. NMR
analysis confirms the presence of Li\textsuperscript{+} ions in both
static and mobile environments. MD simulations of a crystal grain with
adiponitrile-rich boundaries show that boundary ions diffuse 40-90 times
faster than ions inside the grain, so that a boundary population of
5-10\% carries most of the ionic conductance.

\section*{Supplementary Information}

Supplementary information is provided in an available PDF file.

The crystallographic information file (CIF) will be made available upon publication.

\section*{Acknowledgements}

This work is primarily supported by a grant from the National Science
Foundation under award DMR-2138432. Additional support from the
Department of Energy under award DE-SC0023356 is acknowledged.
Crystallographic data was obtained on instrumentation provided by an MRI
grant under award CHE-2215854.

The NMR work at Hunter College was supported in part by the U.S.
National Aeronautics and Space Administration (NASA) \emph{via} the
NASA-Hunter Center for Advanced Energy Storage for Space under
cooperative agreement \#80NSSC24M0177. We thank Dr. Allen Zheng, Hunter
College, for assistance with the PFG-NMR diffusion measurements.

W.A.G. acknowledges~support from the Hong Kong Quantum AI Lab, AIR@
InnoHK of the Hong Kong Government. W.A.G. thanks the~U.S. National
Science Foundation (CBET- 2311117) for~support. The computations
presented here were conducted in the Resnick High Performance Computing
Center, a facility supported by Resnick Sustainability Institute at the
California Institute of Technology.

\section*{Use of Generative AI}

The authors used generative AI tools (ChatGPT 5.2 and Claude Opus 5.5) to
assist in improving the visualization and presentation of results and figures, in transforming the original DOCX manuscript content into TeX format, with the goal of improving the HTML rendering and viewing experience on arXiv, and in optimizing the in-house atomic renderer DRSTA for visualization of
atomic trajectories. No experimentally observed data were generated, altered,
processed, or interpreted using generative AI tools, and no experimental
figures were generated or processed using AI. The use of generative AI was
limited to the purposes described above and followed relevant ethical guidance.

\section*{References}

(1) Joshi, A.; Mishra, D. K.; Singh, R.; Zhang, J.; Ding, Y. A
Comprehensive Review of Solid-State Batteries. \emph{Appl. Energy}
\textbf{2025}, \emph{386}, 125546.
https://doi.org/10.1016/j.apenergy.2025.125546

(2) Piglowska, M.; Kurc, B.; Galinski, M.; Fuc, P.; Kaminska, M.;
Szymlet, N.; Daszkiewicz, P. Challenges for Safe Electrolytes Applied in
Lithium-Ion Cells-A Review. \emph{Materials} \textbf{2021}, \emph{14}
(22), 6783. https://doi.org/10.3390/ma14226783

(3) Han, L.; Lehmann, M. L.; Zhu, J. D.; Liu, T. Y.; Zhou, Z. P.; Tang,
X. M.; Heish, C. T.; Sokolov, A. P.; Cao, P. F.; Chen, X. C.; Saito, T.
Recent Developments and Challenges in Hybrid Solid Electrolytes for
Lithium-Ion Batteries. \emph{Front. Energy Res.} \textbf{2020},
\emph{8}, 202. https://doi.org/10.3389/fenrg.2020.00202

(4) Paul, S. C.; Wunder, S. L.; Zdilla, M. J. Hard--Soft Acid--Base
Interactions Control Ionic Conductivity in Molecular-Crystal-Based
Electrolytes. \emph{ACS Mater. Lett.} \textbf{2026}, \emph{8} (2),
470--475. https://doi.org/10.1021/acsmaterialslett.5c01234

(5) Paul, S. C.; Zdilla, M. J.; Wunder, S. L. Critical Role of Li+···Li+
Distance and Anion Restriction in Conductivity and Lithium-Ion
Transference Number of Molecular Crystals for Solid State Electrolytes.
\emph{Adv. Energy Mater.} \textbf{2025}, \emph{16} (4), e03433.
https://doi.org/10.1002/aenm.202503433

(6) Fall, B.; Sonnenberg, L. A.; Aguirre, J. R.; Paul, S. C.; Prakash,
P.; Venkatnathan, A.; Garaga, M. N.; Goddard, W. A., III; Greenbaum, S.
G.; Zdilla, M. J.; Wunder, S. L. Effect of Anion Mass on Conductivity
and Lithium-Ion Transference Number in the Isomorphic Cocrystals
(Adpn)2LiXF6 (Adpn = Adiponitrile, X = P, As, Sb). \emph{Chem. Mater.}
\textbf{2025}, \emph{37} (5), 1798--1809.
https://doi.org/10.1021/acs.chemmater.4c01374

(7) Paul, S. C.; Goddard, W. A., III; Zdilla, M. J.; Prakash, P.;
Wunder, S. L. Grain Boundary Tuning Determines Iodide and Lithium-Ion
Migration in a Solid Adiponitrile--LiI Molecular Crystal Electrolyte.
\emph{Mater. Horiz.} \textbf{2025}, \emph{12} (24), 10728--10739.
https://doi.org/10.1039/D5MH00900F

(8) Paul, S. C.; Goddard, W. A., III; Zdilla, M. J.; Prakash, P.;
Wunder, S. L. Nanoconfined Grain Boundaries Increase the Conductivity of
Polycrystalline Molecular Crystals. \emph{ACS Mater. Lett.}
\textbf{2026}, \emph{8} (3), 764--771.
https://doi.org/10.1021/acsmaterialslett.5c01267

(9) Zhao, F.; Zhang, S.; Wang, S.; Reid, J. W.; Xia, W.; Liu, J.; King,
G.; Kaduk, J. A.; Liang, J.; Luo, J.; Gao, Y.; Yang, F.; Zhao, Y.; Li,
W.; Alahakoon, S. H.; Guo, J.; Huang, Y.; Sham, T. K.; Mo, Y.; Sun, X.
Anion Sublattice Design Enables Superionic Conductivity in Crystalline
Oxyhalides. \emph{Science} \textbf{2025}, \emph{390} (6769), 199--204.
https://doi.org/10.1126/science.adt9678

(10) Leube, B. T.; Collins, C. M.; Daniels, L. M.; Duff, B. B.; Dang,
Y.; Chen, R.; Gaultois, M. W.; Manning, T. D.; Blanc, F.; Dyer, M. S.;
Claridge, J. B.; Rosseinsky, M. J. Cation Disorder and Large Tetragonal
Supercell Ordering in the Li-Rich Argyrodite Li7Zn0.5SiS6. \emph{Chem.
Mater.} \textbf{2022}, \emph{34} (9), 4073--4087.
https://doi.org/10.1021/acs.chemmater.2c00320

(11) Soulié, J. P.; Renaudin, G.; Černý, R.; Yvon, K. Lithium
Boro-Hydride LiBH4: I. Crystal Structure. \emph{J. Alloys Compd.}
\textbf{2002}, \emph{346}, 200--205.
https://doi.org/10.1016/S0925-8388(02)00521-2

(12) Matsuo, M.; Nakamori, Y.; Orimo, S.; Maekawa, H.; Takamura, H.
Lithium Superionic Conduction in Lithium Borohydride Accompanied by
Structural Transition. \emph{Appl. Phys. Lett.} \textbf{2007}, \emph{91}
(22), 224103. https://doi.org/10.1063/1.2817934

(13) Maekawa, H.; Matsuo, M.; Takamura, H.; Ando, M.; Noda, Y.;
Karahashi, T.; Orimo, S. Halide-Stabilized LiBH4, a Room-Temperature
Lithium Fast-Ion Conductor. \emph{J. Am. Chem. Soc.} \textbf{2009},
\emph{131} (3), 894--895. https://doi.org/10.1021/ja807392k

(14) Matsuo, M.; Takamura, H.; Maekawa, H.; Li, H. W.; Orimo, S.
Stabilization of Lithium Superionic Conduction Phase and Enhancement of
Conductivity of LiBH4 by LiCl Addition. \emph{Appl. Phys. Lett.}
\textbf{2009}, \emph{94} (8), 084103. https://doi.org/10.1063/1.3088857

(15) Prakash, P.; Fall, B.; Aguirre, J.; Sonnenberg, L. A.; Chinnam, P.
R.; Chereddy, S.; Dikin, D. A.; Venkatnathan, A.; Wunder, S. L.; Zdilla,
M. J. A Soft Co-Crystalline Solid Electrolyte for Lithium-Ion Batteries.
\emph{Nat. Mater.} \textbf{2023}, 1--9.
https://doi.org/10.1038/s41563-023-01508-1

(16) Tanaka, K.; Tago, Y.; Kondo, M.; Watanabe, Y.; Nishio, K.;
Hitosugi, T.; Moriya, M. High Li-Ion Conductivity in
Li\{N(SO2F)2\}(NCCH2CH2CN)2 Molecular Crystal. \emph{Nano Lett.}
\textbf{2020}, \emph{20} (11), 8200--8204.
https://doi.org/10.1021/acs.nanolett.0c03313

(17) Katsuragawa, H.; Mori, S.; Tago, Y.; Maeda, S.; Matsuda, S.; Toriu,
H.; Nakayama, R.; Kobayashi, S.; Hitosugi, T.; Moriya, M. Molecular
Crystalline Electrolyte Based on Li\{N(SO2CF3)2\} and Succinonitrile
with Closely Contacted Grain Boundary Interfaces Exhibiting Selective
Li-Ion Conductivity and 5 V-Class Electrochemical Stability. \emph{ACS
Appl. Energy Mater.} \textbf{2025}, \emph{8} (6), 3599--3605.
https://doi.org/10.1021/acsaem.4c03207

(18) Züttel, A.; Borgschulte, A.; Orimo, S.-I. Tetrahydroborates as New
Hydrogen Storage Materials. \emph{Scr. Mater.} \textbf{2007}, \emph{56}
(10), 823--828. https://doi.org/10.1016/j.scriptamat.2007.01.010

(19) Prakash, P.; Shylendran, A.; Fall, B.; Zdilla, M. J.; Wunder, S.
L.; Venkatnathan, A. Mechanism of Ion Conduction and Dynamics in
Tris(N,N-Dimethylformamide) Perchloratosodium Solid Electrolytes.
\emph{J. Phys. Chem. C} \textbf{2022}, \emph{126} (10), 4744--4750.
https://doi.org/10.1021/acs.jpcc.1c09005

(20) Winter, E.; Seipel, P.; Miß, V.; Spannenberger, S.; Roling, B.;
Vogel, M. 7Li NMR Studies of Short-Range and Long-Range Lithium Ion
Dynamics in a Heat-Treated Lithium Iodide-Doped Lithium Thiophosphate
Glass Featuring High Ion Conductivity. \emph{J. Phys. Chem. C}
\textbf{2020}, \emph{124} (52), 28614--28622.
https://doi.org/10.1021/acs.jpcc.0c08801

(21) Böhmer, R.; Jeffrey, K. R.; Vogel, M. Solid-State Li NMR with
Applications to the Translational Dynamics in Ion Conductors.
\emph{Prog. Nucl. Magn. Reson. Spectrosc.} \textbf{2007}, \emph{50}
(2--3), 87--174. https://doi.org/10.1016/j.pnmrs.2006.12.001

(22) Dupke, S.; Langer, T.; Pöttgen, R.; Winter, M.; Passerini, S.;
Eckert, H. Structural Characterization of the Lithium Silicides Li15Si4,
Li13Si4, and Li7Si3 Using Solid State NMR. \emph{Phys. Chem. Chem.
Phys.} \textbf{2012}, \emph{14} (18), 6496--6508.
https://doi.org/10.1039/C2CP24131E

(23) Epp, V.; Gün, Ö.; Deiseroth, H.-J.; Wilkening, M. Highly Mobile
Ions: Low-Temperature NMR Directly Probes Extremely Fast
Li\textsuperscript{+} Hopping in Argyrodite-Type Li\textsubscript{6}
PS\textsubscript{5} Br. \emph{J. Phys. Chem. Lett.} \textbf{2013},
\emph{4} (13), 2118--2123. https://doi.org/10.1021/jz401003a

(24) Sheppard, D.; Xiao, P.; Chemelewski, W.; Johnson, D. D.; Henkelman,
G. A Generalized Solid-State Nudged Elastic Band Method. \emph{J. Chem.
Phys.} \textbf{2012}, \emph{136}, 074103.
https://doi.org/10.1063/1.3684549

(25) Batatia, I.; Kovács, D. P.; Simm, G. N. C.; Ortner, C.; Csányi, G.
MACE: Higher Order Equivariant Message Passing Neural Networks for Fast
and Accurate Force Fields. \emph{Adv. Neural Inf. Process. Syst.}
\textbf{2022}, \emph{35}, 11423--11436.

(26) Batatia, I.; Benner, P.; Chiang, Y.; Elena, A. M.; Kovács, D. P.;
Riebesell, J.; Advincula, X. R.; Asta, M.; Avaylon, M.; Baldwin, W. J.;
Berger, F.; Bernstein, N.; Bhowmik, A.; Bigi, F.; Blau, S. M.; Cărare,
V.; Ceriotti, M.; Chong, S.; Darby, J. P.; De, S.; Della Pia, F.;
Deringer, V. L.; Elijošius, R.; El-Machachi, Z.; Falcioni, F.; Fako, E.;
Ferrari, A. C.; Gardner, J. L. A.; Gawkowski, M. J.; Genreith-Schriever,
A.; George, J.; Goodall, R. E. A.; Grandel, J.; Grey, C. P.; Grigorev,
P.; Han, S.; Handley, W.; Heenen, H. H.; Hermansson, K.; Holm, C.; Ho,
C. H.; Hofmann, S.; Jaafar, J.; Jakob, K. S.; Jung, H.; Kapil, V.;
Kaplan, A. D.; Karimitari, N.; Kermode, J. R.; Kourtis, P.; Kroupa, N.;
Kullgren, J.; Kuner, M. C.; Kuryla, D.; Liepuoniute, G.; Lin, C.;
Margraf, J. T.; Magdău, I.-B.; Michaelides, A.; Moore, J. H.; Naik, A.
A.; Niblett, S. P.; Norwood, S. W.; O'Neill, N.; Ortner, C.; Persson, K.
A.; Reuter, K.; Rosen, A. S.; Rosset, L. A. M.; Schaaf, L. L.; Schran,
C.; Shi, B. X.; Sivonxay, E.; Stenczel, T. K.; Svahn, V.; Sutton, C.;
Swinburne, T. D.; Tilly, J.; van der Oord, C.; Vargas, S.;
Varga-Umbrich, E.; Vegge, T.; Vondrák, M.; Wang, Y.; Witt, W. C.; Wolf,
T.; Zills, F.; Csányi, G. A Foundation Model for Atomistic Materials
Chemistry. \emph{arXiv} \textbf{2023}, 2401.00096.

\clearpage
\let\addcontentsline\savedaddcontentsline
\setcounter{page}{1}
\renewcommand{\thepage}{S\arabic{page}}
\setcounter{section}{0}
\setcounter{subsection}{0}

\begin{center}
{\Large\bfseries Supporting Information\par}
\vspace{0.6em}
{\large\bfseries A Solid-Solid Phase Transition Reverses Carrier Selectivity in the Molecular Crystal Electrolyte (Adpn)$_2$LiI\par}
\vspace{1.0em}
Shujit Chandra Paul$^{1}$, Giselle de Araujo$^{2}$, Phillip Stallworth$^{2}$, Steven Greenbaum$^{2}$,\\
William A. Goddard III$^{3}$, Prabhat Prakash$^{3,*}$, Stephanie Wunder$^{1,*}$, Michael Zdilla$^{1,*}$\\[0.8em]
{\small $^{1}$Department of Chemistry, Temple University, Philadelphia, PA 19122, USA\\
$^{2}$Department of Physics, Hunter College, CUNY, New York, New York 10065, USA\\
$^{3}$Materials and Process Simulation Center (MSC), California Institute of Technology, Pasadena, California 91125, USA\\[0.5em]
$^{*}$Corresponding authors: Prabhat Prakash (\href{mailto:pprakash@caltech.edu}{\texttt{pprakash@caltech.edu}});\\
Stephanie Wunder (\href{mailto:slwunder@temple.edu}{\texttt{slwunder@temple.edu}}); Michael Zdilla (\href{mailto:mzdilla@temple.edu}{\texttt{mzdilla@temple.edu}})}
\end{center}
\vspace{1em}
\tableofcontents
\clearpage

\section{Experimental details
}\label{experimental-details}

\subsection{Materials and Synthesis}

The adiponitrile and anhydrous LiI were purchased from Sigma Aldrich.
The adiponitrile was distilled and degassed in a Schlenk line before
placing in the glove box. The molecular crystal was made by heating the
LiI at 160 °C in the required amount of adiponitrile to make Adpn:LiI
samples in ratios of \emph{n}:1 where \emph{n} = 2, 2.4, 3, which are
henceforth labelled as (Adpn)\emph{n}LiI, even though these represent
heterogeneous mixtures of (Adpn)2LiI and excess Adpn in the grain
boundary.

\subsection{Characterization
}\label{characterization}

Single-Crystal X-ray Diffraction (SCXRD) was obtained using Mo-Ka
radiation on a Bruker D8 Venture diffractometer. The high temperature
phase was solved by putting the crystal in epoxy with hardener and then
mount the crystal on the top of a glass capillary. Differential scanning
calorimetry (DSC) on a TA Instruments Model 250 DSC was used to
determine melt temperatures (T\textsubscript{m}), crystallization
temperatures (T\textsubscript{c}) and glass transition temperatures
(T\textsubscript{g}) at a scan rate of 10 °C/min (unless specified
otherwise) under ultra-pure N\textsubscript{2} purge. Raman spectra were
recorded in the 2000--2500\,cm\textsuperscript{-1} region at room
temperature with a Horiba LabRAM HR Evolution Raman spectrometer, using
an excitation wavelength of 532nm and a grating groove density of 600
grooves/millimeter. The temperature dependent Raman was conducted using
smart substrates (VAHEAT, Interherence, Germany) with a controlled
heating and cooling rate. Conductivity measurements were obtained with
2032-type coin cell and impedance data was measured by AC
electrochemical impedance spectroscopy (EIS), acquired with a Gamry
Interface 1000 potentiostat/galvanostat/ZRA (zero-resistance-ammeter) in
the frequency range 0.1-1MHz synchronized with an ESPEC benchtop chamber
(Model-BTU-133) in a temperature range between 0\,°C and 110\,°C with
increments of 10\,°C. Li\textsuperscript{+} ion transference numbers
(t\textsubscript{Li}\textsuperscript{+}) were obtained by the method of
Vincent and Bruce\textsuperscript{1}. The
Li\textsuperscript{0}/Electrolyte/Li\textsuperscript{0} coin cell was
made by melting the molecular crystal onto a glass fiber separator. The
impedance was measured before and after DC polarization, and the
transference number was calculated using:

\(t_{Li}^{+} = \frac{I_{s}\left( \mathrm{\Delta}V - I_{0}R_{0} \right)}{I_{0}\left( \mathrm{\Delta}V - I_{s}R_{s} \right)}\)
(3)

where \ensuremath{\Delta}V is the applied voltage, I\textsubscript{s} and
R\textsubscript{s} are the steady-state current and resistance,
respectively, and I\textsubscript{0} and R\textsubscript{0} are the
initial current and resistance, respectively.

\subsection{NMR details
}\label{nmr-details}

\textsuperscript{7}Li NMR measurements were performed on a Bruker Avance
400 spectrometer operating at a magnetic field strength of 9.4 T
(\textsuperscript{7}Li resonance frequency of 155 MHz). Sample
temperature was controlled with a variable-temperature unit and
regulated compressed air gas flow, and samples were allowed to
equilibrate for 15 min prior to acquisition. Static
\textsuperscript{7}Li spectra were acquired using a hard pulse (pulse
width 3 \ensuremath{\mu}s) with a 5 s recycle delay, 1600 scans and 2048 data points.
Spin--lattice relaxation times (T\textsubscript{1}) were measured at 50 °C and 85 °C
using the inversion-recovery pulse sequence, and raw data were acquired
with a list of 16 delay (\ensuremath{\tau}) values. T\textsubscript{1} were then
obtained by fitting the data using the
\emph{T\textsubscript{1}/T\textsubscript{2}} module in \emph{TopSpin}.
\textsuperscript{7}Li magic angle spinning (MAS) NMR spectra were
collected on a standard 4 mm Bruker MAS probe, spinning at 10 kHz, with
2048 scans, a 5 s recycle delay, and 32,768 data points. All chemical
shifts were referenced to 1.0 M LiCl aqueous solution at 0.00 ppm,
measured under identical conditions. Pulsed Field Gradient (PFG) NMR was
employed to investigate the self-diffusion coefficients of
Li\textsuperscript{+}~ions in (Adpn)\textsubscript{2}LiI samples. The samples were
transferred to 5 mm glass NMR tubes inside an argon-filled glovebox and
flame-sealed to prevent air and moisture exposure. Diffusion experiments
were carried out at 10\,°C using a Fuji Electric Co. model PXF4
temperature controller. Measurements were conducted on a 300 MHz NMR
spectrometer (7T, Agilent/Varian DDR console) equipped with a Doty
Z-spec PFG probe. The diffusion time (\ensuremath{\Delta}) and gradient pulse duration (\ensuremath{\delta})
were optimized between 10 ms and 2 ms, respectively. The maximum
gradient strength applied was 590 and 720 G·cm\textsuperscript{-}\textsuperscript{1}. Self-diffusion
coefficients were extracted using the Stejskal-Tanner\textsuperscript{2}
equation:

\begin{longtable}[]{@{}
  >{\raggedright\arraybackslash}p{(\columnwidth - 2\tabcolsep) * \real{0.6102}}
  >{\raggedright\arraybackslash}p{(\columnwidth - 2\tabcolsep) * \real{0.3898}}@{}}
\toprule\noalign{}
\begin{minipage}[b]{\linewidth}\raggedright
\[\frac{\text{I}}{I_{\text{0}}}\text{= }\exp\left( \text{-}{\text{D}\text{\ensuremath{\gamma}}}^{\text{2}}\text{g}^{\text{2}}\text{\ensuremath{\delta}}^{\text{2}}\left( \text{\ensuremath{\Delta} - }\frac{\text{\ensuremath{\delta}}}{\text{3}} \right) \right)\text{ }\]
\end{minipage} & \begin{minipage}[b]{\linewidth}\raggedright
(1)
\end{minipage} \\
\midrule\noalign{}
\endhead
\bottomrule\noalign{}
\endlastfoot
\end{longtable}

Here, \emph{I} is the signal intensity with an applied gradient,
\emph{I\textsubscript{0}} is the initial signal intensity, \emph{\ensuremath{\gamma}} is
the gyromagnetic ratio of the investigated nucleus, \emph{g} is the
gradient strength applied, and D is the self-diffusion coefficient.

\subsection{Computational details
}\label{computational-details}

A unit cell of HT Adpn\textsubscript{2}LiI structure (12.619 x 12.619 x
5.219 Å\textsuperscript{3}, 68 atoms) was optimized with the DFT-D3
framework as implemented in the Quantum ESPRESSO (QE) v7.2
software\textsuperscript{3}. The projector augmented-wave (PAW) method
was employed to describe the electron--ion interactions, as implemented
in QE. Exchange--correlation effects were treated within the
Perdew--Burke--Ernzerhof (PBE) generalized gradient approximation (GGA)
functional, augmented with Grimme's D3 dispersion correction (with
Becke--Johnson damping) to account for long-range van der Waals
interactions. The plane-wave basis set was used with a kinetic energy
cutoff of 816.3 eV for wavefunctions and 3265.4 eV for charge density,
ensuring convergence of total energy and forces. Brillouin zone
integrations were carried out using a Monkhorst--Pack k-point mesh of
1x1x2. Structural relaxations were performed until the forces on atoms
were less than 0.026 eV/Å. All pseudopotentials were taken from the
PSLibrary PAW dataset compatible with QE 7.2.

The minimum energy path (MEP) for the conduction of
Li\textsuperscript{+} and I\textsuperscript{-} ions was calculated using
the nudged elastic band (NEB) module in QE. Since the
\emph{c}-crystallographic direction exhibits the shortest distance
channels for both Li\textsuperscript{+} and I\textsuperscript{-} ions,
we examined the activation energy (E\textsubscript{a}) barrier along
this direction. To mimic the reactant and product geometries for ion hop
events, first a 1 x 1 x 3 supercell (12.619 x 12.619 x 15.658
Å\textsuperscript{3}, 204 atoms) was created and optimized with a
similar relaxation protocol as described above. Further, reactant and
product pairs of supercells were created with the removal of a single
Li\textsuperscript{+} (and I\textsuperscript{-}, in a separate supercell
pair) ion from each supercell. For each Li\textsuperscript{+}~and
I\textsuperscript{-} image pairs, NEB calculations were performed with
the DFT-D3 framework while tracing 9 intermediate images in between. The
MEP was assumed to be converged once the norm of the forces orthogonal
to the path reached \ensuremath{\le} 0.03 eV/Å.

The conversion of the low-temperature phase of Adpn\textsubscript{2}LiI
to the high-temperature Adpn\textsubscript{2}LiI was simulated using
classical molecular dynamics (MD) with a modified class-II
non-polarizable force-field. Several gas-phase finite structures of
Adpn\textsubscript{2}LiI (at low-temperature) were used to determine the
partial atomic charges on Li\textsuperscript{+}~and I\textsuperscript{-}
ions (+ 0.705 and -- 0.705, respectively). These charges were calculated
from the gas-phase DFT calculations (\ensuremath{\omega}B97x-D4/def2-tzvp) and charges
from electrostatic potential (CHELPG)\textsuperscript{4} method using
Orca 6.1.0 software\textsuperscript{5}. Bonded and non-bonded parameters
for Adpn were adapted from elsewhere.\textsuperscript{6} A supercell
(20400 atoms) was constructed with 3x5x5 unit cells of the
low-temperature (LT) phase structure (272 atoms). The atomistic
trajectory to observe conversion from the LT to the high-temperature
(HT) phase was obtained by simulating this supercell in a
velocity-Verlet integrator with a 1 fs timestep, velocity-rescale
thermostat (100 fs time constant) and C-rescale barostat (isotropic, 1
ps time constant). The system was first preequilibrated from 100 K to
300 K with a heating rate of 200 K/ns, followed by 1 ns equilibration at
300 K. Continuing in \emph{NpT} ensemble conditions, the temperature of
the system was further gradually increased with a heating rate of 100
K/ns from 300 K to 600 K, while equilibrating for 1 ns at each 100 K
breakpoint (at 400 K and 500 K). A slower heating rate of 20
K-ns\textsuperscript{-1} was used to reassess the structural
transformations between 300 to 400 K. A cutoff distance of 12 Å was used
for the calculation of Coulombic and vdW (Lennard-Jones) forces. To
further model the phase transformation, a 1x2x2 supercell of LT
structure of Adpn\textsubscript{2}LiI was simulated with a pull of 2000
kJ-mol\textsuperscript{-1}-nm\textsuperscript{-1} was used to glide
alternate layers of Li\textsuperscript{+} and I\textsuperscript{-} ions
along \emph{z}-direction of the box. All MD simulations were performed
with Gromacs 2024.4 software package\textsuperscript{7}.

\textbf{Solid-state NEB of the phase transformation.} The LT to HT
transformation path was computed with the solid-state nudged elastic
band (SS-NEB) method,\textsuperscript{8} in which the cell vectors are
optimized together with the atomic positions (TSASE with
ASE\textsuperscript{9}). Energies, forces and stresses were obtained
from a MACE\textsuperscript{10} machine-learned interatomic potential
(MLIP) fine-tuned from the MACE-MP-0 foundation
model\textsuperscript{11} on 3,008
PBE\textsuperscript{12}+D3(BJ)\textsuperscript{13} configurations of
(Adpn)\textsubscript{2}LiI (272-atom cells, 600 eV cutoff, 1\ensuremath{\times}2\ensuremath{\times}2
k-points) using VASP.\textsuperscript{14} The endpoints were the
DFT-optimized LT and HT structures (8 formula units, 272 atoms). A
concerted path in one LT cell (11 images) was started from an
interpolation of the molecular internal coordinates refined with the
image-dependent pair potential,\textsuperscript{15} and optimized with
FIRE\textsuperscript{16} and a climbing image.\textsuperscript{17} VASP
single points on all images of the final band were used to validate the
MLIP. To test sequential conversion, the all structures of the band
images were replicated to a 2\ensuremath{\times}2\ensuremath{\times}2 supercell (2,176 atoms, 64 formula
units) in which the four rods along c convert one after another (17
images). Energies are reported per formula unit of the simulation cell.

Workflow design and visualization was done using DRSTA-0.1.0
package.\textsuperscript{18}

\section{Additional Crystal data
}\label{additional-crystal-data}

\includegraphics[width=\textwidth,height=0.82\textheight,keepaspectratio]{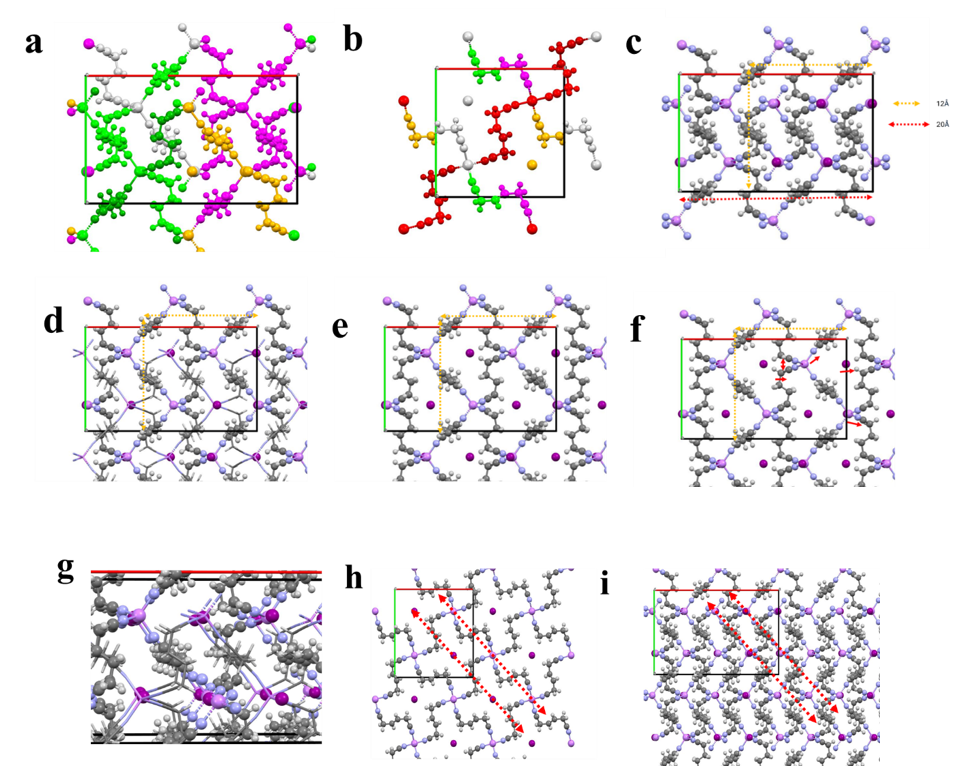}

\textbf{Figure S1:} Transformation of LT to HT phase for the 2:1
Adpn:LiI sample.

\section{Additional DSC data
}\label{additional-dsc-data}

With 5\% LiPF\textsubscript{6} doping, the solid--solid phase transition moves from
about 35 °C in the undoped crystal to about 40 °C. After melting, the
doped sample takes much longer to switch phases during the heating cycle
around 30 minutes instead of \textasciitilde8 minutes for the undoped
one. This longer window let us quickly cool the coin cell back to room
temperature and measure conductivity without crossing a phase boundary.

\begin{center}\includegraphics[width=\textwidth,height=0.68\textheight,keepaspectratio]{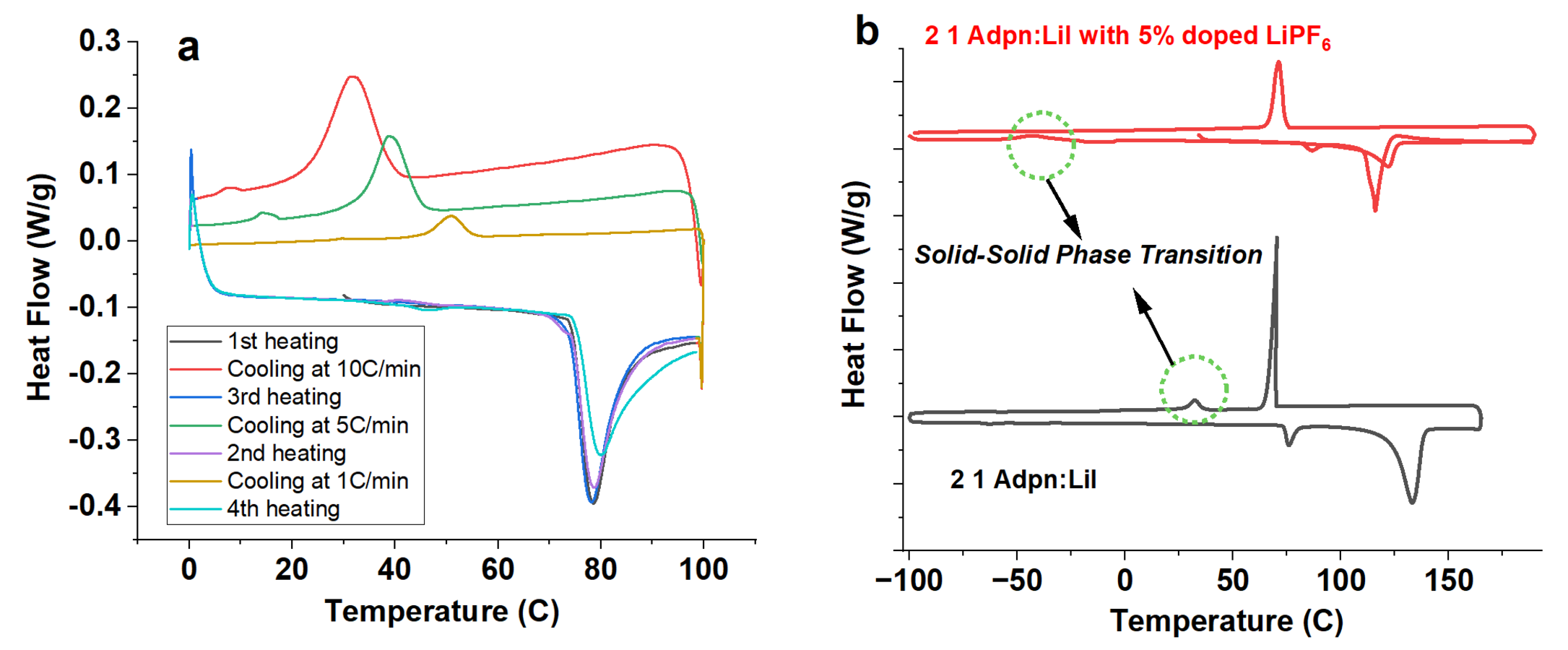}\end{center}

\textbf{Figure S2:} (\textbf{a}) Heating and cooling rate dependent DSC
data for the 2:1 Adpn:LiI sample for the solid-solid phase transition;
(\textbf{b}) DSC data of 2:1 Adpn:LiI and 2:1 Adpn:LiI doped with
5\% 2:1 Adpn:LiPF\textsubscript{6} at 10 \textsuperscript{0}C/min
heating/cooling rates. Note: transformation to low temperature phase
occurred during the heating cycle.

\section{Temperature Dependent Transference
Number}\label{temperature-dependent-transference-number}

\includegraphics[width=\textwidth,height=0.82\textheight,keepaspectratio]{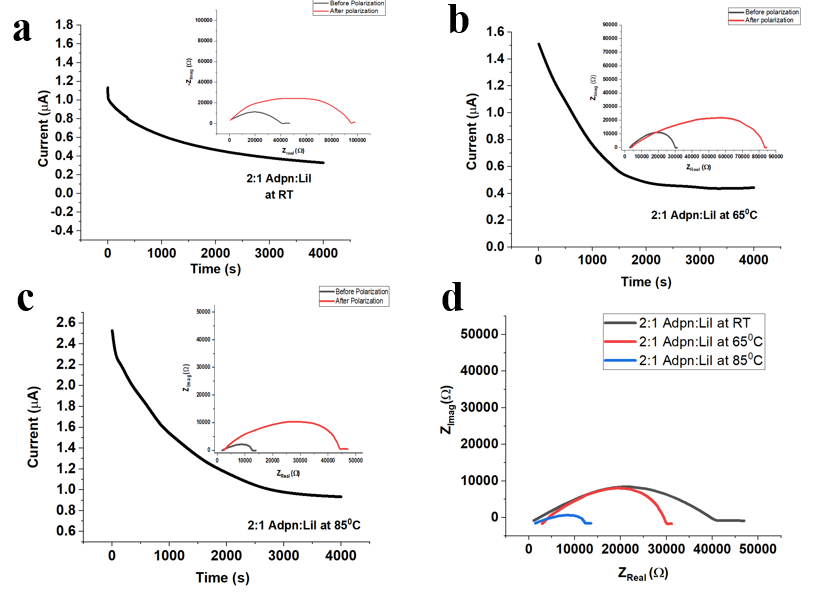}

\textbf{Figure S3:} (a-c) Temperature dependent transference data for
the 2:1 Adpn:LiI sample and (d) comparison of interfacial resistance as
a function of temperature and phase.

\section{Cyclic Voltammetry data}\label{cyclic-voltammetry-data}

\includegraphics[width=\textwidth,height=0.82\textheight,keepaspectratio]{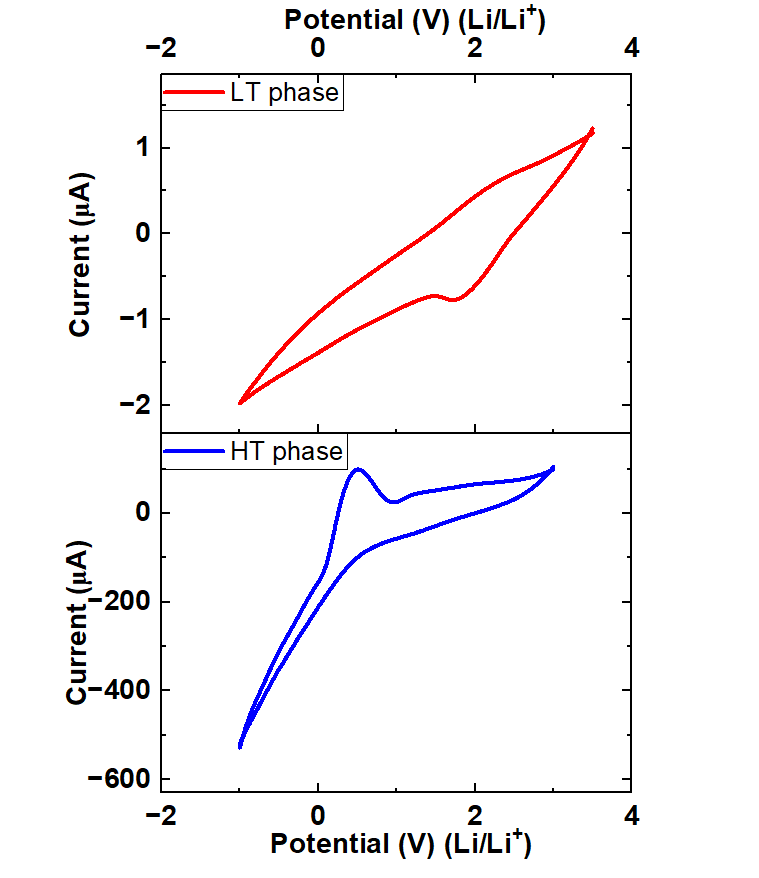}

\textbf{Figure S4:} Temperature dependent CV data for the 2:1 Adpn:LiI
sample

\section{Additional NMR data}\label{additional-nmr-data}

We performed \textsuperscript{7}Li MAS to obtain more specific
information on the lithium environments (Figure S5c). Under 10 kHz MAS,
the \textsuperscript{7}Li central transition at 50 °C (red) appears as a
structured, multi-component profile, with two resolved features
separated by 0.29 ppm on the low-field side (Signal 1) and a broad peak
(FWHM \ensuremath{\approx} 200 Hz) on the high field side (Signal 2), evidencing the
coexistance of a broad distribution of Li environments (Figure 3e). At
80 °C (blue), the separation of the low-field features reduced to only
0.14, while the main high-field peak narrows (FWHM \ensuremath{\approx} 150 Hz). This
behavior reflects faster site-to-site hopping and a reduction in EFG
heterogeneity at higher temperature, consistent with the solid--solid
transition observed in the static spectra. Although most Li sites merge
into a dynamically averaged environment, a small residual shoulder
persists on the low-field side at 80 °C. This feature indicates that the
minority population of Li ions remains in less symmetric or
defect-related environments, which are not fully averaged out by the
increased mobility.

\includegraphics[width=\textwidth,height=0.82\textheight,keepaspectratio]{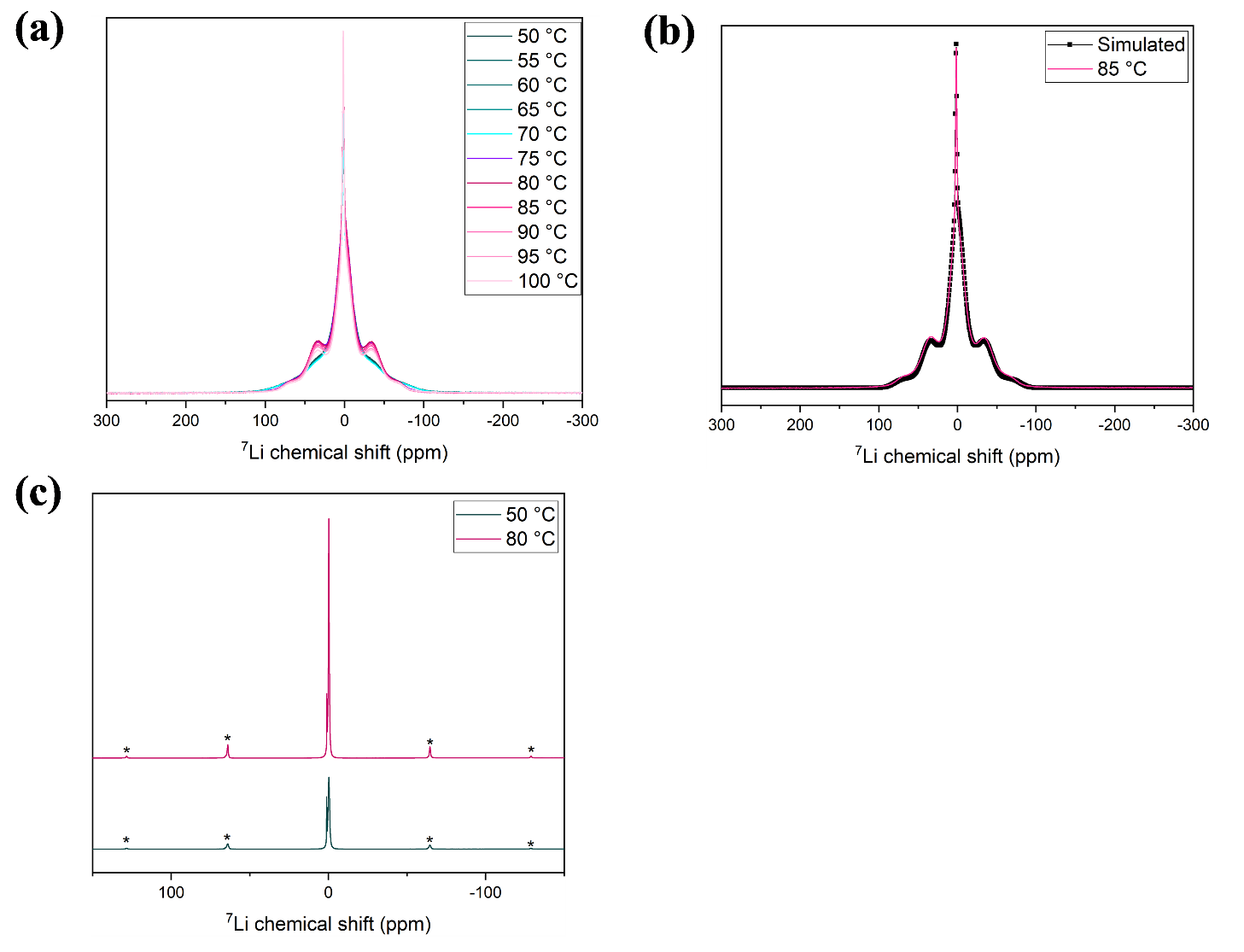}

\textbf{Figure S5.} \textbf{(a)} Static \textsuperscript{7}Li NMR
spectra of (Adpn)\textsubscript{2}LiI measured from 50 to 100 °C. \textbf{(b)}
Experimental and simulated \textsuperscript{7}Li spectrum at 85 °C
\textbf{(c)} \textsuperscript{7}Li MAS NMR spectra acquired at 10 kHz
spinning speed at 50 °C and 80 °C.

\section{Additional Simulation data
}\label{additional-simulation-data}

\textbf{Simulation of phase transformation}

Molecular dynamics (MD) simulations with classical interatomic
potentials were used to simulate the structural transformation from LT
to HT phase in Adpn\textsubscript{2}LiI, starting with a 3x5x5 supercell
from the LT-Adpn\textsubscript{2}LiI unit cell. was heated at two
heating rates: i) 100 K-ns\textsuperscript{-1} to heat from 300 K to 600
K (provided in \textbf{Figure S7}), to identify transformations and ii)
20 K-ns\textsuperscript{-1} to heat from 300 K to 400 K, with 1ns
equilibration at each 20 K intermediate temperature point.
\textbf{Figure 5a} shows that the mass density, which dropped gradually
until 7 ns and 360 K, dropped dramatically in the 7.6 -- 7.9 ns window
during the 360 -- 380 K temperature ramp. The calculated non-bonded
interaction energy (NBIE) profiles for Adpn and I\textsuperscript{-}
around Li\textsuperscript{+} during this thermal transformation show
significant change \textbf{(Figure 5b)}. A Li\textsuperscript{+} ion in
these molecular crystals coordinates with four CN groups from four Adpn
molecules, with an average NBIE of -250 kJ-mol\textsuperscript{-1}
(\textasciitilde62.5 kJ-mol\textsuperscript{-1} per Li-N non-covalent
interaction).

Since the I\textsuperscript{-} ion is in its own anionic vdW space, with
a minimum distance of 5.4 Å from a Li\textsuperscript{+} ion, the
average Li\textsuperscript{+}···I\textsuperscript{-} NBIE is -12
kJ-mol\textsuperscript{-1}. As the crystal structure collapses in the
360 -- 380 K window, Li\textsuperscript{+}···I\textsuperscript{-}
ion-pairs phase-separate (with NBIE increased negatively to
\textasciitilde{} -100 kJ-mol\textsuperscript{-1}), while having
\textasciitilde{} two-three Adpn interacting with the
Li\textsuperscript{+}···I\textsuperscript{-}···Li\textsuperscript{+}
linear wires. The trajectory of Li\textsuperscript{+~}ions at 300 -- 340
K (\textbf{Figure 5c, d)} shows only the cage vibrations over a period
of 1ns. For the 340 -- 360 K temperature window, some
Li\textsuperscript{+~}ions can be seen moving in the regular vacancy
sites, formed via their mutual exchange with
I\textsuperscript{-}~occupancy sites (\textbf{Figure 5e)}, suggesting a
short-lived LT -- HT transformation, before the structure is seen
collapsed in the 360 -- 380 K window (\textbf{Figure 5f}).

\subsection{Adiponitrile conformers
}\label{adiponitrile-conformers}

\includegraphics[width=\textwidth,height=0.82\textheight,keepaspectratio]{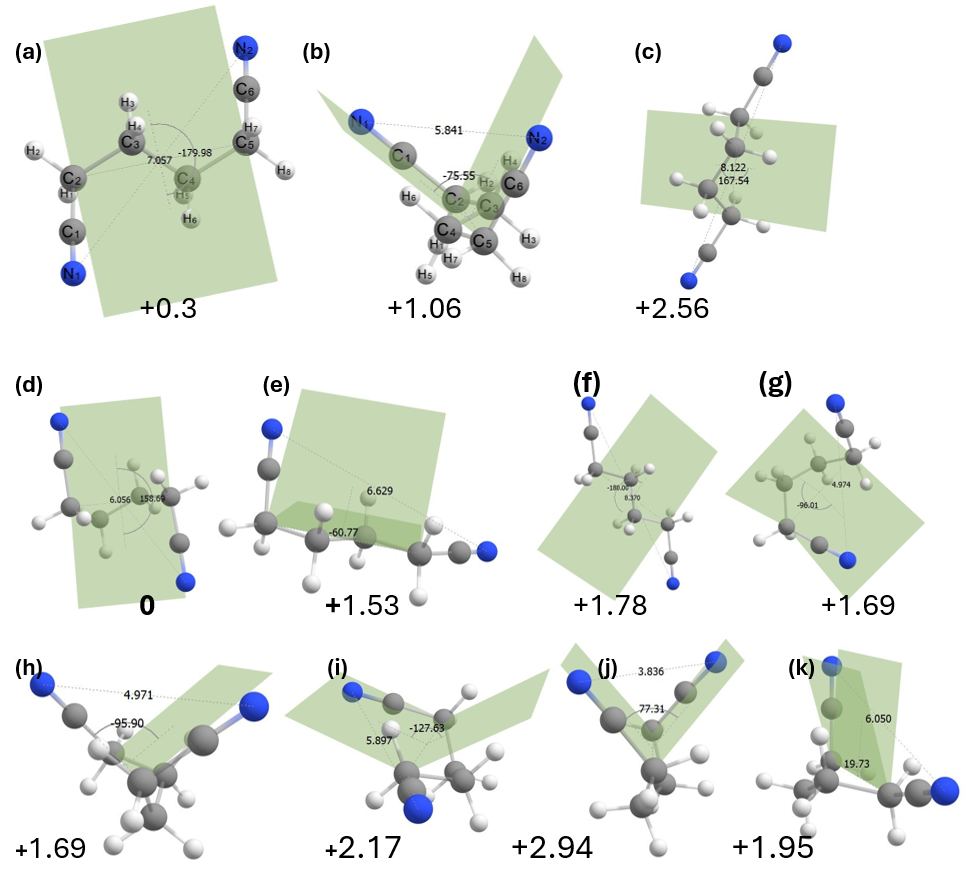}

\textbf{Figure S6.} Gas-phase (\ensuremath{\omega}B97x-D4 def2-tzvpp) optimized geometries
of two conformers of \textbf{(a)} trans C1-C2-C5-C6 conformer (present
50\% in LT, 100 \% in HT) and \textbf{(b)} cis C2-C2-C5-C6 conformer
(present 50 \% in LT). Dotted lines between N1 and N2 atoms show
end-to-end distance, green plane shows the plane of dihedral, other
dotted lines between C2 and C5 show the shared bond of C1-C2-C5-C6
dihedral. \textbf{(c)} -- \textbf{(k)} other nine low-energy separated
conformers, with \textbf{(d)} being the global minima in gas-phase.
Bottom text on each subpanel shows the relative free energies
(\ensuremath{\Delta}G\textsubscript{298}) in kcal/mol, computed from DFT + vibrational
frequency calculations.

\subsection{Temperature, density and NBIE during simulated annealing
from 100 K to 600
K}\label{temperature-density-and-nbie-during-simulated-annealing-from-100-k-to-600-k}

\includegraphics[width=\textwidth,height=0.82\textheight,keepaspectratio]{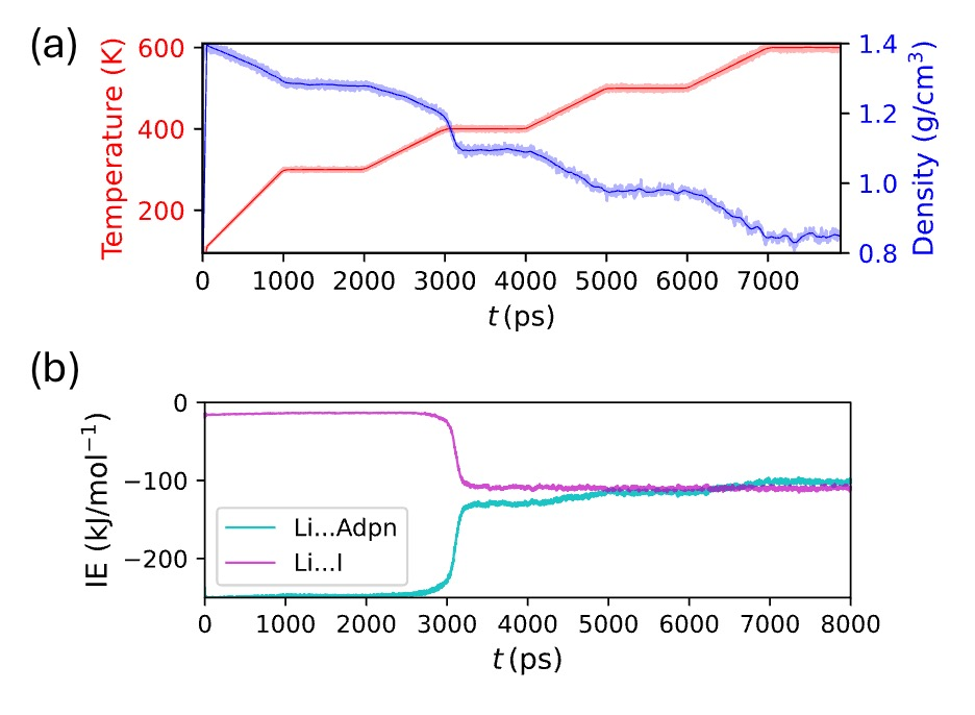}

\textbf{Figure S7. (a)} Temperature and density profiles for the
simulated heating (in NPT ensemble conditions) of
LT-Adpn\textsubscript{2}LiI structure from 100 K to 600 K, with first
ramping to 300 K in 1 ns and then heating rates of 100 K/ns including
intermediate equilibrations at three temperature points: 300, 400, and
500 K. \textbf{(b)} Li\textsuperscript{+} \ldots Adpn and
Li\textsuperscript{+}\ldots{} I\textsuperscript{-} non-bonded
interaction energies (NBIE), normalized for one Li\textsuperscript{+}
ion in the heating-equilibration trajectory shown in \emph{a}.

\subsection{Temperature, density and NBIE during simulated annealing
from 300 K to 400
K}\label{temperature-density-and-nbie-during-simulated-annealing-from-300-k-to-400-k}

\includegraphics[width=\textwidth,height=0.82\textheight,keepaspectratio]{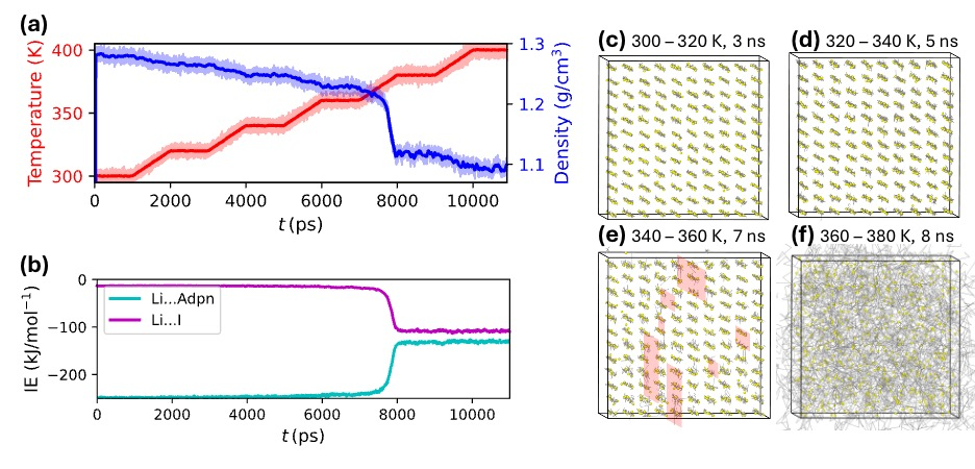}

\textbf{Figure S8. (a)} Temperature and density profiles for the
simulated heating (in NPT ensemble conditions) of the LT Adpn2LiI
structure from 300 K to 400 K, with heating rates of 20 K/ns including
intermediate equilibrations at four temperature points: 320, 340, 360
and 380 K. \textbf{(b)} Li+ \ldots Adpn and Li+\ldots{} I- non-bonded
interaction energies (IE), normalized for one Li+ ion in the
heating-equilibration trajectory shown in a. \textbf{(c) -- (f)}
Snapshots of Li+ ion (yellow) position and trajectory shows: only cage
vibrations (c and d) in LT phase at 300 -- 340 K; significant movement
of Li+ ions between channels suggests the transition phase (LT $\rightarrow$ HT, 340
-- 360 K, e); decomposition of solvent matrix (\textgreater{} 360 K, f).
The density profile in a and NBIE in b also shows the sudden drop and
reversal of preferred interactions for Li+ in the same corresponding
temperature window (360 -- 380 K).

\subsection{Temperature, density and NBIE at NpT simulations at 360
K}\label{temperature-density-and-nbie-at-npt-simulations-at-360-k}

A 10 ns continuation at 360 K shows that the crystal structure collapses
within 2 ns \textbf{(Figure S9)}, indicating that the LT $\rightarrow$ HT
transformation window of temperature is quite narrow (as also seen in
experiments).

\includegraphics[width=\textwidth,height=0.82\textheight,keepaspectratio]{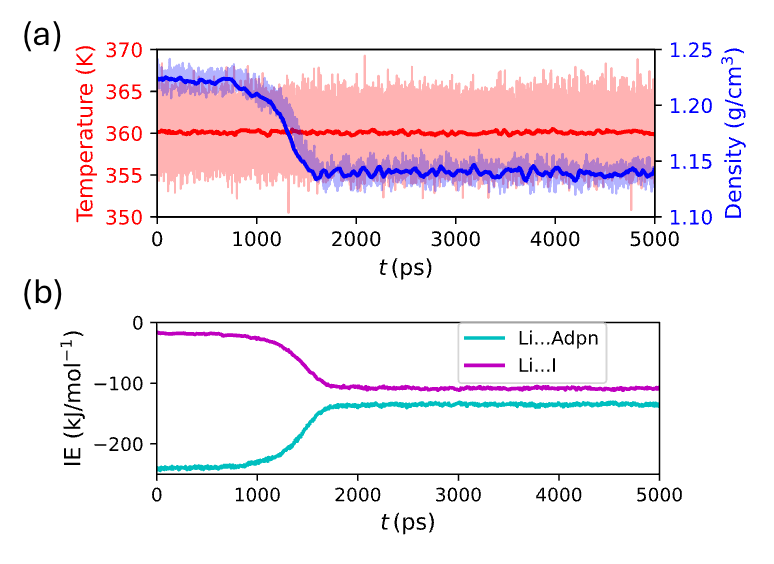}

\textbf{Figure S9. (a)} Temperature and density profiles for
LT-Adpn\textsubscript{2}LiI structure at 360 K under NpT ensemble
conditions, \textbf{(b)} Li\textsuperscript{+} \ldots Adpn and
Li\textsuperscript{+}\ldots{} I\textsuperscript{-} non-bonded
interaction energies (NBIE), normalized for one Li\textsuperscript{+}
ion in the heating-equilibration trajectory shown in \emph{a}.

\subsection{Distribution histogram for the diffusion
coefficients of Li\textsuperscript{+}
cations}\label{distribution-histogram-for-the-diffusion-coefficients-of-li-cations}

Even at 360 K, before the structure could collapse in standalone
simulations, here for 1ns, the integrity of the framework remains intact
with several more Li\textsuperscript{+} ions seen in the gray regions
suggesting more ionic reorganization to locally facilitate LT $\rightarrow$ HT.
However, the transition doesn't last further, as at 380 K (and at 400
K), the network collapses to form
Li\textsuperscript{+}···I\textsuperscript{-}···Li\textsuperscript{+}
chains and clusters with \ensuremath{\le} 3Adpn average coordination, facilitating a
superionic diffusion (10\textsuperscript{-10} -- 10\textsuperscript{-9}
m\textsuperscript{2}-s\textsuperscript{-1}, \textbf{Figure 11e, f}) as
expected in the decomposed state.

\includegraphics[width=\textwidth,height=0.82\textheight,keepaspectratio]{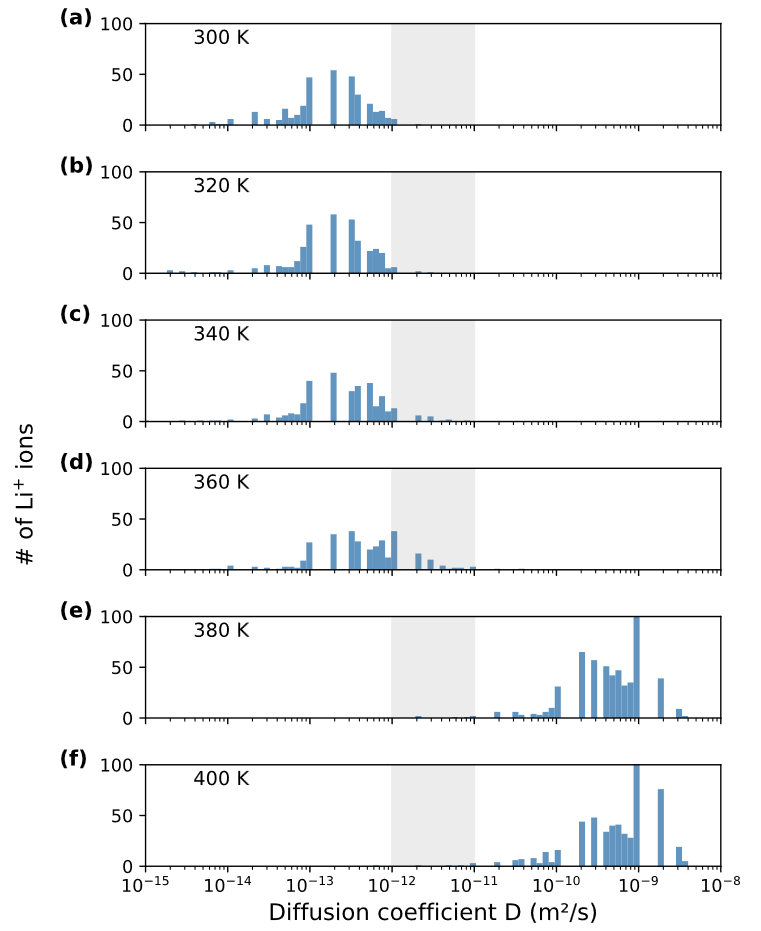}

\textbf{Figure S10.} Distribution histogram for the diffusion
coefficients of Li+ ions at (a) 300 K, (b) 320 K, (c) 340 K, (d) 360 K,
(e) 380 K, (f) 400 K. Gray area shows the charge carriers corresponding
to HT phase.

\subsection{Distribution histogram for the diffusion
coefficients of I\textsuperscript{-}
anions}\label{distribution-histogram-for-the-diffusion-coefficients-of-i--anions}

\includegraphics[width=\textwidth,height=0.82\textheight,keepaspectratio]{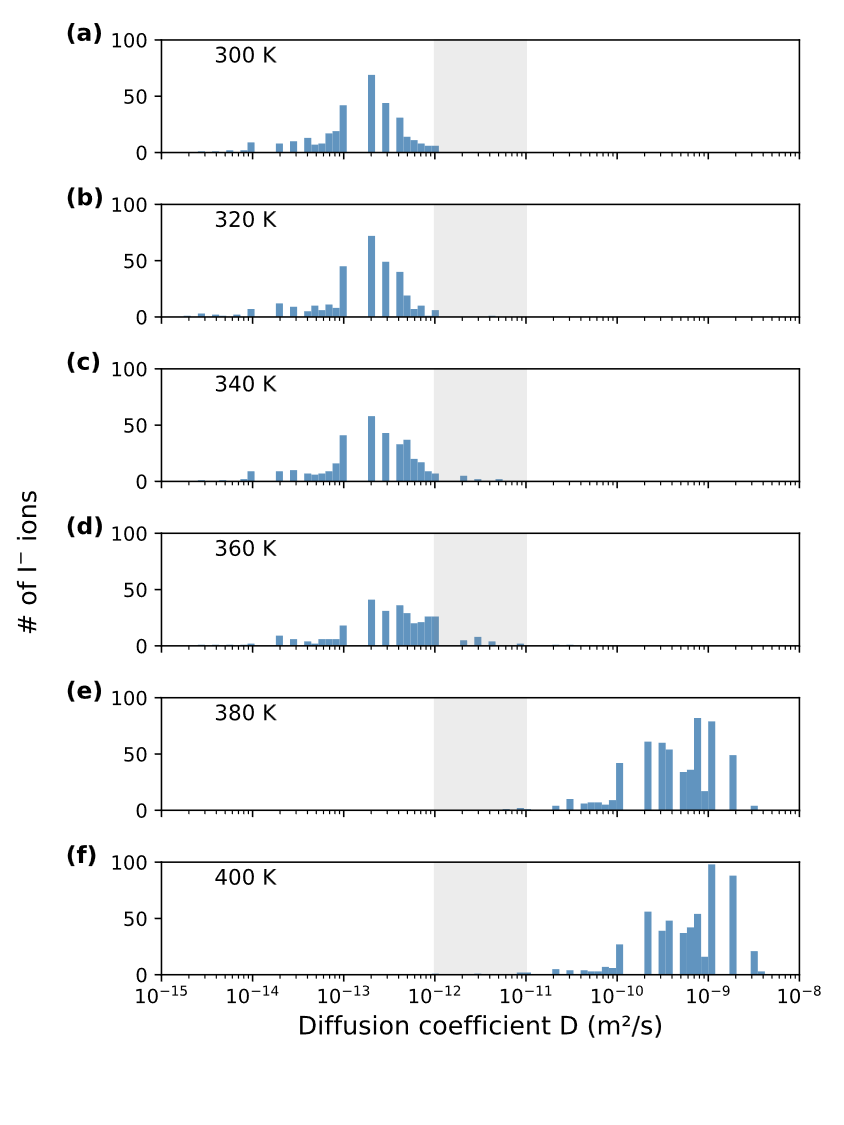}

\textbf{Figure S11.} Distribution histogram for the diffusion
coefficients of I- ions at (a) 300 K, (b) 320 K, (c) 340 K, (d) 360 K,
(e) 380 K, (f) 400 K. Gray area shows the charge carriers corresponding
to HT phase.

\textbf{Solid-state NEB of the LT \ensuremath{\rightarrow} HT transformation}

The MACE-MP-0 (medium) foundation model\textsuperscript{10,11} with the
D3(BJ) dispersion correction\textsuperscript{13} does not reproduce the
DFT reaction energy or forces well enough for this transformation (E(HT)
- E(LT) = +62 meV/f.u. against +5 meV/f.u. from DFT; force RMSE 158--180
meV Å\textsuperscript{-1}. It was therefore fine-tuned on existing
PBE\textsuperscript{12}+D3(BJ)\textsuperscript{13} calculations of
(Adpn)\textsubscript{2}LiI (Table S1): 272-atom cells, 600 eV plane-wave
cutoff, with stresses recorded at every ionic step. The training
settings are listed in \textbf{Table S2}; validation curves and errors
are shown in \textbf{Figure S12} and \textbf{Table S3}.

\textbf{Table S1.} DFT training and test data. Every second ionic step
of the relaxations and every fifth step of the NEB were kept and frames
with an atomic force above 10 eV Å\textsuperscript{-1} were removed.

\begin{longtable}[]{@{}
  >{\raggedright\arraybackslash}p{(\columnwidth - 8\tabcolsep) * \real{0.3548}}
  >{\raggedright\arraybackslash}p{(\columnwidth - 8\tabcolsep) * \real{0.0968}}
  >{\raggedright\arraybackslash}p{(\columnwidth - 8\tabcolsep) * \real{0.1183}}
  >{\raggedright\arraybackslash}p{(\columnwidth - 8\tabcolsep) * \real{0.1183}}
  >{\raggedright\arraybackslash}p{(\columnwidth - 8\tabcolsep) * \real{0.3118}}@{}}
\toprule\noalign{}
\begin{minipage}[b]{\linewidth}\raggedright
\textbf{Source}
\end{minipage} & \begin{minipage}[b]{\linewidth}\raggedright
\textbf{k-points}
\end{minipage} & \begin{minipage}[b]{\linewidth}\raggedright
\textbf{Ionic steps}
\end{minipage} & \begin{minipage}[b]{\linewidth}\raggedright
\textbf{Frames kept}
\end{minipage} & \begin{minipage}[b]{\linewidth}\raggedright
\textbf{Use}
\end{minipage} \\
\midrule\noalign{}
\endhead
\bottomrule\noalign{}
\endlastfoot
Variable-cell relaxations of 11 images of an initial MACE-MP-0 path &
1\ensuremath{\times}2\ensuremath{\times}2 & 4,044 & 2,030 & energies, forces, stresses; image 5 held out \\
Relaxations of the LT and HT endpoints & 1\ensuremath{\times}2\ensuremath{\times}2 & 171 & 87 & energies,
forces, stresses \\
\ensuremath{\Gamma}-point VASP SS-NEB (9 images, 500 steps, not converged) & \ensuremath{\Gamma} & 4,500 &
891 & forces and stresses only; image 5 held out \\
Total & & 8,715 & 3,008 & 2,701 training / 307 test \\
\end{longtable}

\textbf{Table S2.} Fine-tuning settings for replication.

\begin{longtable}[]{@{}
  >{\raggedright\arraybackslash}p{(\columnwidth - 2\tabcolsep) * \real{0.2473}}
  >{\raggedright\arraybackslash}p{(\columnwidth - 2\tabcolsep) * \real{0.7527}}@{}}
\toprule\noalign{}
\begin{minipage}[b]{\linewidth}\raggedright
\textbf{Setting}
\end{minipage} & \begin{minipage}[b]{\linewidth}\raggedright
\textbf{Value}
\end{minipage} \\
\midrule\noalign{}
\endhead
\bottomrule\noalign{}
\endlastfoot
Foundation model & MACE-MP-0 medium (2023-12-03), fine-tuned without
replay of the foundation data \\
Targets & VASP energies, forces and stresses minus D3(BJ) (PBE
parameters) \\
Loss & Huber; weights: energy 1, forces 10, stress 100 \\
Optimization & learning rate 10\textsuperscript{-3} with plateau
scheduler (patience 5), EMA 0.99, batch size 4, 200 epochs \\
Precision, hardware & float64; one NVIDIA H100 GPU, 9.6 h \\
Validation / test & random 5\% of the training frames / held-out image-5
trajectories \\
\end{longtable}

\textbf{Table S3.} Errors on the held-out image-5 trajectories, which
were not used in training. Energy RMSEs are after removing each
model\textquotesingle s mean offset per trajectory.

\begin{longtable}[]{@{}
  >{\raggedright\arraybackslash}p{(\columnwidth - 4\tabcolsep) * \real{0.4839}}
  >{\raggedright\arraybackslash}p{(\columnwidth - 4\tabcolsep) * \real{0.2581}}
  >{\raggedright\arraybackslash}p{(\columnwidth - 4\tabcolsep) * \real{0.2581}}@{}}
\toprule\noalign{}
\begin{minipage}[b]{\linewidth}\raggedright
\textbf{Held-out trajectory, quantity}
\end{minipage} & \begin{minipage}[b]{\linewidth}\raggedright
\textbf{MACE-MP-0 + D3}
\end{minipage} & \begin{minipage}[b]{\linewidth}\raggedright
\textbf{Fine-tuned + D3}
\end{minipage} \\
\midrule\noalign{}
\endhead
\bottomrule\noalign{}
\endlastfoot
Relaxation: energy RMSE (meV/f.u.) & 88.5 & 3.1 \\
Relaxation: force RMSE (meV Å\textsuperscript{-1}) & 158 & 15.4 \\
Relaxation: stress RMSE (GPa) & 0.273 & 0.019 \\
\ensuremath{\Gamma}-point NEB: energy RMSE (meV/f.u.) & 121 & 11.3 \\
\ensuremath{\Gamma}-point NEB: force RMSE (meV Å\textsuperscript{-1}) & 180 & 26.5 \\
\ensuremath{\Gamma}-point NEB: stress RMSE (GPa) & 0.286 & 0.023 \\
\end{longtable}

\includegraphics[width=\textwidth,height=0.82\textheight,keepaspectratio]{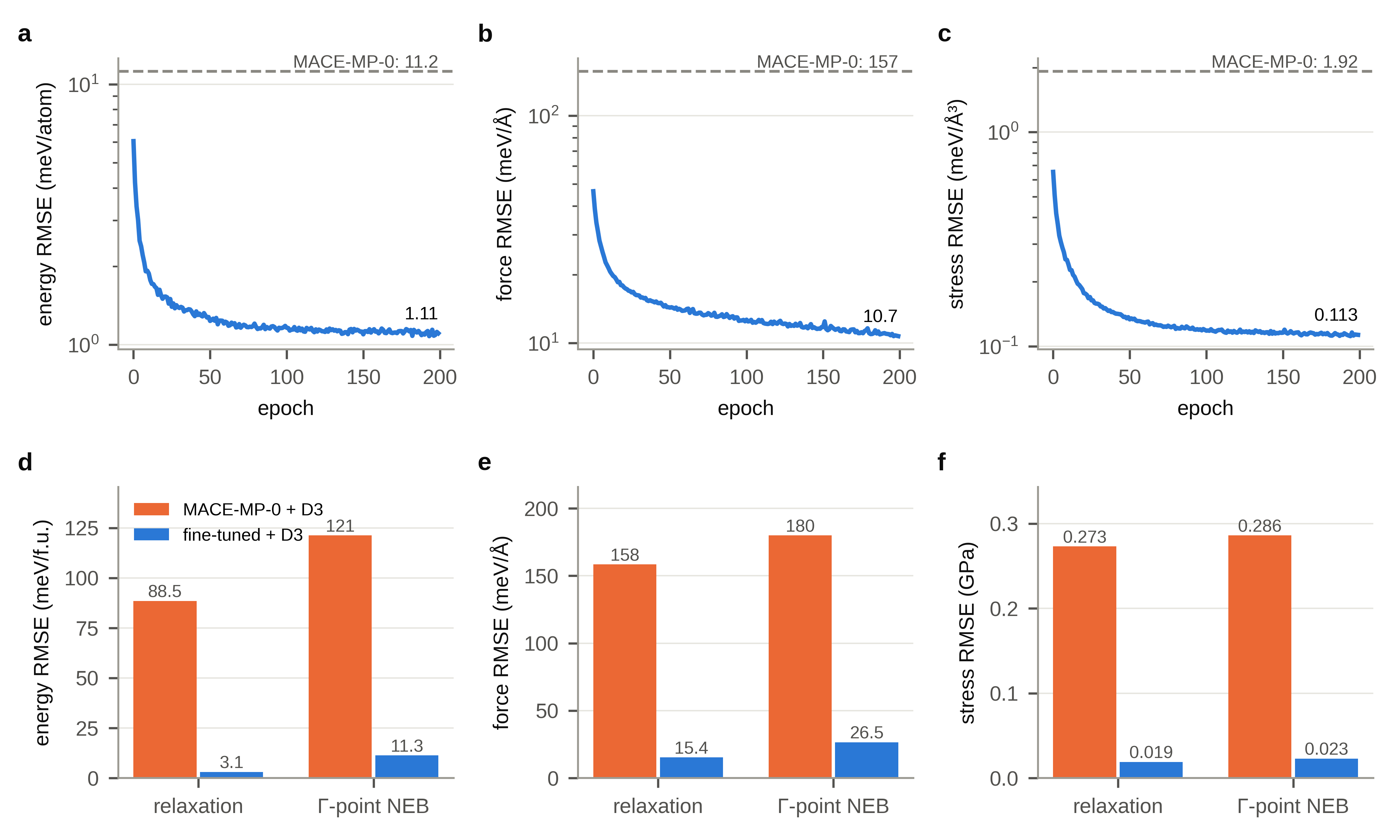}

\textbf{Figure S12.} Fine-tuning of the MLIP. (a--c) Validation RMSE of
energy, force and stress per epoch (log scale); dashed lines, MACE-MP-0
before fine-tuning. (d, e, f) Errors on the held-out image-5
trajectories for MACE-MP-0 + D3 and the fine-tuned model + D3.

\subsection{Concerted path in one LT
cell}\label{concerted-path-in-one-lt-cell}

\textbf{Endpoints.} The DFT-optimized LT and HT structures (V = 3,005.7
and 3,068.0 Å\textsuperscript{3}) are minima of the fine-tuned MLIP
(largest atomic force \textasciitilde0.01 eV Å\textsuperscript{-1},
stress \textasciitilde0) and were used unchanged. The MLIP gives E(HT) -
E(LT) = +11.1 meV/f.u., against +4.9 meV/f.u. from DFT \textbf{(Table
S5)}.

\textbf{Path.} The SS-NEB\textsuperscript{8} was run with TSASE and
ASE\textsuperscript{9} using 11 images, FIRE,\textsuperscript{16} a
maximum step of 0.02 Å and a cell weight of 2: 1,500 iterations without
and then up to 3,700 iterations with a climbing
image.\textsuperscript{17} The starting band was built by interpolating
the molecular internal coordinates (Z-matrix, with spherical
interpolation of the molecular orientations) and refining it with the
image-dependent pair potential;\textsuperscript{15} a band started from
the unconverged \ensuremath{\Gamma}-point DFT NEB diverged, because its strongly expanded
images lie outside the training data. The runs were stopped and
discarded if any interatomic distance fell below 0.9 Å or a cell angle
deviated by more than 5-10° from 90°. With the default FIRE settings
momentum built up and the band ran away after reaching the largest
atomic force of \textasciitilde0.1 eV Å\textsuperscript{-1}; the last
band before the runaway (climbing-image iteration 3,700; largest atomic
force 0.12 eV Å\textsuperscript{-1}) is reported \textbf{(Figure S13,
Table S4)}. The mid-path cells are 7-15\% larger than the endpoints and
tilt by up to 5°.

\textbf{Table S4.} The concerted NE-band (climbing-image iteration
3,700). E, energy relative to LT (meV/f.u.); V, cell volume; tilt,
largest deviation of a cell angle from 90°; F\textsubscript{max},
largest atomic force; F RMSE, force RMSE of the MLIP against DFT.

\begin{longtable}[]{@{}
  >{\raggedright\arraybackslash}p{(\columnwidth - 22\tabcolsep) * \real{0.2050}}
  >{\raggedright\arraybackslash}p{(\columnwidth - 22\tabcolsep) * \real{0.0723}}
  >{\raggedright\arraybackslash}p{(\columnwidth - 22\tabcolsep) * \real{0.0723}}
  >{\raggedright\arraybackslash}p{(\columnwidth - 22\tabcolsep) * \real{0.0723}}
  >{\raggedright\arraybackslash}p{(\columnwidth - 22\tabcolsep) * \real{0.0723}}
  >{\raggedright\arraybackslash}p{(\columnwidth - 22\tabcolsep) * \real{0.0723}}
  >{\raggedright\arraybackslash}p{(\columnwidth - 22\tabcolsep) * \real{0.0723}}
  >{\raggedright\arraybackslash}p{(\columnwidth - 22\tabcolsep) * \real{0.0723}}
  >{\raggedright\arraybackslash}p{(\columnwidth - 22\tabcolsep) * \real{0.0723}}
  >{\raggedright\arraybackslash}p{(\columnwidth - 22\tabcolsep) * \real{0.0723}}
  >{\raggedright\arraybackslash}p{(\columnwidth - 22\tabcolsep) * \real{0.0723}}
  >{\raggedright\arraybackslash}p{(\columnwidth - 22\tabcolsep) * \real{0.0723}}@{}}
\toprule\noalign{}
\begin{minipage}[b]{\linewidth}\raggedright
\textbf{Image}
\end{minipage} & \begin{minipage}[b]{\linewidth}\raggedright
\textbf{LT}
\end{minipage} & \begin{minipage}[b]{\linewidth}\raggedright
\textbf{1}
\end{minipage} & \begin{minipage}[b]{\linewidth}\raggedright
\textbf{2}
\end{minipage} & \begin{minipage}[b]{\linewidth}\raggedright
\textbf{3}
\end{minipage} & \begin{minipage}[b]{\linewidth}\raggedright
\textbf{4}
\end{minipage} & \begin{minipage}[b]{\linewidth}\raggedright
\textbf{5}
\end{minipage} & \begin{minipage}[b]{\linewidth}\raggedright
\textbf{6}
\end{minipage} & \begin{minipage}[b]{\linewidth}\raggedright
\textbf{7}
\end{minipage} & \begin{minipage}[b]{\linewidth}\raggedright
\textbf{8}
\end{minipage} & \begin{minipage}[b]{\linewidth}\raggedright
\textbf{9}
\end{minipage} & \begin{minipage}[b]{\linewidth}\raggedright
\textbf{HT}
\end{minipage} \\
\midrule\noalign{}
\endhead
\bottomrule\noalign{}
\endlastfoot
E, MLIP & 0 & 519 & 620 & 753 & 982 & 834 & 798 & 919 & 681 & 441 &
11 \\
E, DFT & 0 & 479 & 591 & 716 & 956 & 855 & 784 & 886 & 658 & 416 & 9 \\
V (Å\textsuperscript{3}) & 3006 & 3285 & 3281 & 3299 & 3288 & 3325 &
3359 & 3461 & 3301 & 3236 & 3068 \\
tilt (°) & 0.0 & 1.3 & 0.2 & 2.0 & 3.6 & 1.4 & 3.5 & 4.9 & 4.7 & 3.0 &
0.0 \\
F\textsubscript{max}, MLIP (eV Å\textsuperscript{-1}) & 0.01 & 0.09 &
0.03 & 0.03 & 0.12 & 0.03 & 0.10 & 0.08 & 0.05 & 0.11 & 0.01 \\
F\textsubscript{max}, DFT (eV Å\textsuperscript{-1}) & 0.01 & 0.15 &
0.12 & 0.11 & 0.14 & 0.12 & 0.12 & 0.10 & 0.09 & 0.13 & 0.01 \\
F RMSE (meV Å\textsuperscript{-1}) & 2 & 14 & 16 & 16 & 23 & 23 & 19 &
17 & 14 & 11 & 1 \\
\end{longtable}

\textbf{Table S5.} Energy of HT relative to LT at 0 K.

\begin{longtable}[]{@{}
  >{\raggedright\arraybackslash}p{(\columnwidth - 2\tabcolsep) * \real{0.6774}}
  >{\raggedright\arraybackslash}p{(\columnwidth - 2\tabcolsep) * \real{0.3226}}@{}}
\toprule\noalign{}
\begin{minipage}[b]{\linewidth}\raggedright
\textbf{Method}
\end{minipage} & \begin{minipage}[b]{\linewidth}\raggedright
\textbf{E(HT) - E(LT) (meV/f.u.)}
\end{minipage} \\
\midrule\noalign{}
\endhead
\bottomrule\noalign{}
\endlastfoot
VASP PBE+D3(BJ), DFT-optimized endpoints & +4.9 \\
VASP PBE+D3(BJ) single points at the endpoints of the band & +9.0 \\
Fine-tuned MLIP + D3(BJ) & +11.1 \\
MACE-MP-0 + D3(BJ) & +62.2 \\
Experiment: \ensuremath{\Delta}H\textsubscript{ss} from DSC (for comparison) & \ensuremath{\approx} +49 \\
\end{longtable}

\includegraphics[width=\textwidth,height=0.82\textheight,keepaspectratio]{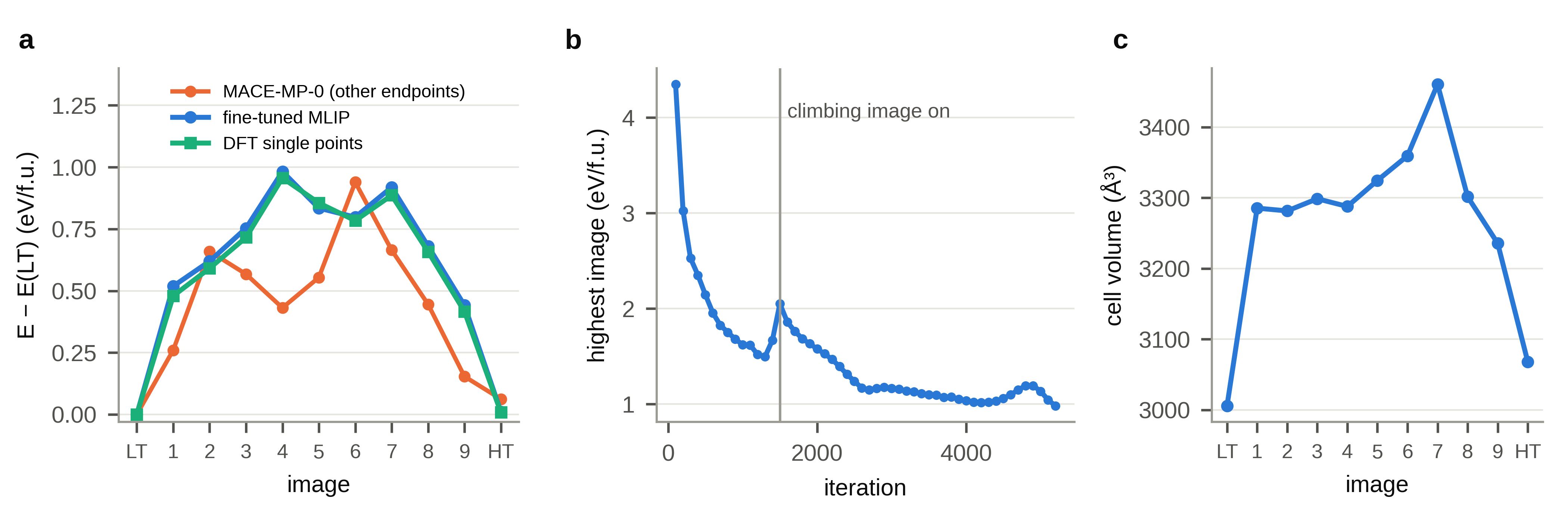}

\textbf{Figure S13.} Concerted SS-NEB in one LT cell. (a) Energy along
the NEB from the fine-tuned MLIP and from DFT single points on the same
images, together with the final NEB of an earlier, unconverged run with
MACE-MP-0 (endpoints relaxed with MACE-MP-0). (b) Highest image energy
during the run; the climbing image was switched on at iteration 1,500.
(c) Cell volume along the NEB path.

\includegraphics[width=\textwidth,height=0.82\textheight,keepaspectratio]{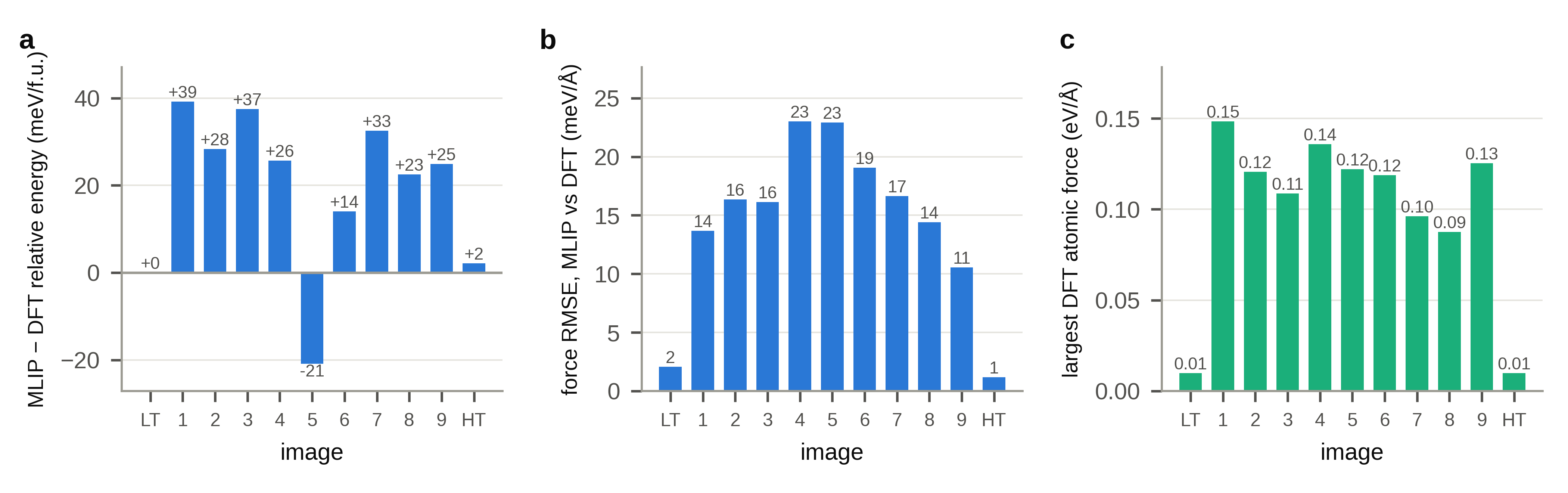}

\textbf{Figure S14.} DFT check of the concerted band, per image: (a)
MLIP - DFT energy relative to LT, (b) force RMSE of the MLIP against
DFT, (c) largest DFT atomic force.

\subsection{Ion columns and Adpn conformers along the
path}\label{ion-columns-and-adpn-conformers-along-the-path}

Of the eight Li\textsuperscript{+}--I\textsuperscript{-} columns along c
in one LT cell, four (one Li\textsuperscript{+} and one
I\textsuperscript{-} per cell each) slide by ±c/2 (6.23 Å; two by +c/2
and two by -c/2), so that their Li\textsuperscript{+} and
I\textsuperscript{-} interchange each other\textquotesingle s LT sites.
The other four columns move by at most 0.56 Å (Li\textsuperscript{+})
and 1.78 Å (I\textsuperscript{-}). The 16 Adpn molecules move by only
\textasciitilde0.2 Å for their COM. The 8 gauche molecules (C1-C2-C5-C6
dihedral 83° in LT) rotate to trans (180°), and the 8 trans molecules
remain trans (their dihedral passes through 126-180°). Along the NEB,
two iodide columns complete their slide by image 4 and the other two by
image 9, the gauche molecules rotate between images 2 and 10, and the Li
columns move mainly between images 8 and 10 \textbf{(Figure 5b)}.

\subsection{Rod-by-rod path in a 2\ensuremath{\times}2\ensuremath{\times}2
supercell}\label{rod-by-rod-path-in-a-222-supercell}

\textbf{Construction.} The validated NEB images were used to build a
2\ensuremath{\times}2\ensuremath{\times}2 supercell (2,176 atoms, 64 f.u.) of eight blocks, each one LT
cell, in which every block can take the geometry of any image of the
concerted band on a common supercell lattice. Molecules were kept whole
and assigned to the block containing their COM. The conversion progress
of a block, used for the coloring in \textbf{Figures 5c} and
\textbf{S16}, is d\textsubscript{LT}/(d\textsubscript{LT} +
d\textsubscript{HT}), where d\textsubscript{LT} and d\textsubscript{HT}
are the RMS distances of its atoms from the LT and HT references; it is
0 for LT and 1 for HT. Because an ion column runs through the two blocks
stacked along c, a column can slide only once both have converted.

\textbf{Table S6.} Rod-path states S\textsubscript{n} of the 2\ensuremath{\times}2\ensuremath{\times}2
supercell, in which n of the 8 blocks (one LT cell each) have converted.
Relaxed: the state relaxed on its own; on the band: the same state after
the SS-NEB.

\begin{longtable}[]{@{}
  >{\raggedright\arraybackslash}p{(\columnwidth - 10\tabcolsep) * \real{0.0860}}
  >{\raggedright\arraybackslash}p{(\columnwidth - 10\tabcolsep) * \real{0.1613}}
  >{\raggedright\arraybackslash}p{(\columnwidth - 10\tabcolsep) * \real{0.2796}}
  >{\raggedright\arraybackslash}p{(\columnwidth - 10\tabcolsep) * \real{0.1613}}
  >{\raggedright\arraybackslash}p{(\columnwidth - 10\tabcolsep) * \real{0.1505}}
  >{\raggedright\arraybackslash}p{(\columnwidth - 10\tabcolsep) * \real{0.1613}}@{}}
\toprule\noalign{}
\begin{minipage}[b]{\linewidth}\raggedright
\textbf{State}
\end{minipage} & \begin{minipage}[b]{\linewidth}\raggedright
\textbf{Converted blocks (rods)}
\end{minipage} & \begin{minipage}[b]{\linewidth}\raggedright
\textbf{Description}
\end{minipage} & \begin{minipage}[b]{\linewidth}\raggedright
\textbf{Relaxed (eV per supercell)}
\end{minipage} & \begin{minipage}[b]{\linewidth}\raggedright
\textbf{Relaxed (meV/f.u.)}
\end{minipage} & \begin{minipage}[b]{\linewidth}\raggedright
\textbf{On the band (meV/f.u.)}
\end{minipage} \\
\midrule\noalign{}
\endhead
\bottomrule\noalign{}
\endlastfoot
S\textsubscript{0} & 0 (0) & all LT & 0.0 & 0 & 0 \\
S\textsubscript{2} & 2 (1) & one HT rod in LT & 28.2 & 440 & 384 \\
S\textsubscript{4} & 4 (2) & HT slab, half of the cell & 44.4 & 693 &
544 \\
S\textsubscript{6} & 6 (3) & one LT rod in HT & 27.2 & 426 & 350 \\
S\textsubscript{8} & 8 (4) & all HT & 0.7 & 11 & 11 \\
\end{longtable}

\textbf{Table S7.} Rod-by-rod band (climbing-image iteration 2,375). Rod
columns give the conversion progress of each rod (mean of its two
blocks; 0 = LT, 1 = HT).

\begin{longtable}[]{@{}
  >{\raggedright\arraybackslash}p{(\columnwidth - 14\tabcolsep) * \real{0.0860}}
  >{\raggedright\arraybackslash}p{(\columnwidth - 14\tabcolsep) * \real{0.1828}}
  >{\raggedright\arraybackslash}p{(\columnwidth - 14\tabcolsep) * \real{0.1828}}
  >{\raggedright\arraybackslash}p{(\columnwidth - 14\tabcolsep) * \real{0.1398}}
  >{\raggedright\arraybackslash}p{(\columnwidth - 14\tabcolsep) * \real{0.1022}}
  >{\raggedright\arraybackslash}p{(\columnwidth - 14\tabcolsep) * \real{0.1022}}
  >{\raggedright\arraybackslash}p{(\columnwidth - 14\tabcolsep) * \real{0.1022}}
  >{\raggedright\arraybackslash}p{(\columnwidth - 14\tabcolsep) * \real{0.1022}}@{}}
\toprule\noalign{}
\begin{minipage}[b]{\linewidth}\raggedright
\textbf{Image}
\end{minipage} & \begin{minipage}[b]{\linewidth}\raggedright
\textbf{State / converting}
\end{minipage} & \begin{minipage}[b]{\linewidth}\raggedright
\textbf{E (eV per supercell)}
\end{minipage} & \begin{minipage}[b]{\linewidth}\raggedright
\textbf{E (meV/f.u.)}
\end{minipage} & \begin{minipage}[b]{\linewidth}\raggedright
\textbf{Rod 1}
\end{minipage} & \begin{minipage}[b]{\linewidth}\raggedright
\textbf{Rod 2}
\end{minipage} & \begin{minipage}[b]{\linewidth}\raggedright
\textbf{Rod 3}
\end{minipage} & \begin{minipage}[b]{\linewidth}\raggedright
\textbf{Rod 4}
\end{minipage} \\
\midrule\noalign{}
\endhead
\bottomrule\noalign{}
\endlastfoot
0 & S\textsubscript{0} & 0.0 & 0 & 0.00 & 0.00 & 0.00 & 0.00 \\
1 & rod 1 & 20.8 & 324 & 0.32 & 0.06 & 0.11 & 0.07 \\
2 & rod 1 & 31.6 & 493 & 0.50 & 0.08 & 0.14 & 0.08 \\
3 & rod 1 & 25.7 & 401 & 0.64 & 0.08 & 0.16 & 0.08 \\
4 & S\textsubscript{2} & 24.6 & 384 & 0.74 & 0.14 & 0.18 & 0.07 \\
5 & rod 2 & 42.0 & 657 & 0.75 & 0.38 & 0.18 & 0.12 \\
6 & rod 2 & 43.4 & 678 & 0.76 & 0.55 & 0.18 & 0.13 \\
7 & rod 2 & 43.4 & 678 & 0.80 & 0.68 & 0.18 & 0.15 \\
8 & S\textsubscript{4} & 34.8 & 544 & 0.78 & 0.78 & 0.18 & 0.18 \\
9 & rod 3 & 39.4 & 616 & 0.83 & 0.83 & 0.30 & 0.20 \\
10 & rod 3 & 43.2 & 675 & 0.84 & 0.82 & 0.45 & 0.21 \\
11 & rod 3 & 32.9 & 514 & 0.88 & 0.83 & 0.64 & 0.23 \\
12 & S\textsubscript{6} & 22.4 & 350 & 0.90 & 0.83 & 0.80 & 0.25 \\
13 & rod 4 & 28.3 & 441 & 0.93 & 0.84 & 0.83 & 0.33 \\
14 & rod 4 & 30.6 & 478 & 0.94 & 0.84 & 0.86 & 0.48 \\
15 & rod 4 & 14.7 & 230 & 0.94 & 0.88 & 0.93 & 0.68 \\
16 & S\textsubscript{8} & 0.7 & 11 & 1.00 & 1.00 & 1.00 & 1.00 \\
\end{longtable}

\includegraphics[width=\textwidth,height=0.82\textheight,keepaspectratio]{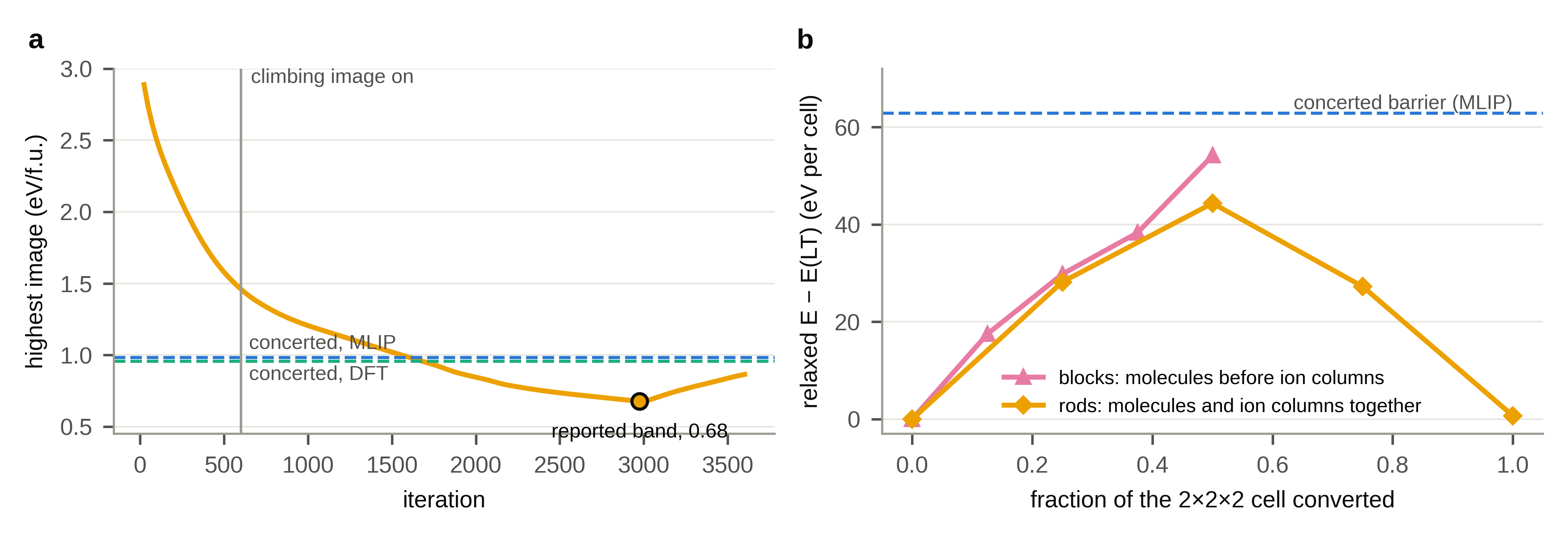}

\textbf{Figure S15.} 2\ensuremath{\times}2\ensuremath{\times}2 supercell. (a) Highest image of the
rod-by-rod band during the SS-NEB (climbing image from iteration 600),
with the concerted barriers (MLIP and DFT) for reference; the circle
marks the reported band. (b) Relaxed intermediate states of the block
route (molecules before ion columns) and the rod route (molecules and
ion columns together); dashed line, concerted MLIP barrier scaled to the
supercell.

\includegraphics[width=\textwidth,height=0.82\textheight,keepaspectratio]{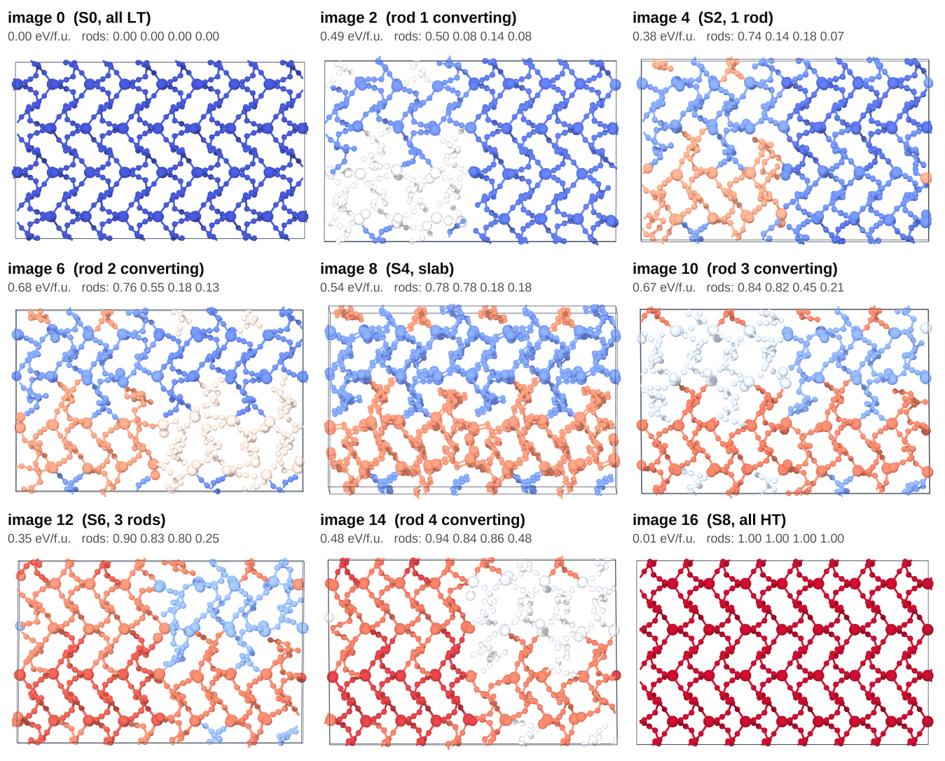}

\textbf{Figure S16.} Views down c of every second image of the
rod-by-rod band (hydrogens hidden), with atoms colored by the conversion
progress of their LT cell (blue, LT; white, half-way; red, HT). Panel
titles give the image, the state, the energy in eV/f.u. and the progress
of the four rods.

\subsection{Limitations}\label{limitations}

\begin{enumerate}
\def\labelenumi{(\roman{enumi})}
\item
  The 2\ensuremath{\times}2\ensuremath{\times}2 band is not converged and its barrier is an estimate between
  the slab state (0.54 eV/f.u.) and 0.68 eV/f.u.
\item
  The MLIP was trained on homogeneous LT, HT and concerted-path
  structures, whereas the 2\ensuremath{\times}2\ensuremath{\times}2 energies are dominated by LT/HT
  interfaces, which have not yet been checked against DFT.
\item
  The concerted barrier refers to a path in which all 8 formula units of
  the cell transform together and is an upper bound for the crystal
\item
  All energies are static 0 K energies without zero-point or thermal
  contributions.
\end{enumerate}

\subsection{Grain boundary models and
simulations}\label{grain-boundary-models-and-simulations}

The 3\ensuremath{\times}5\ensuremath{\times}5 LT supercell (600 Li\textsuperscript{+}, 600
I\textsuperscript{-}, 1200 Adpn; 20,400 atoms; \textasciitilde61 \ensuremath{\times} 62 \ensuremath{\times}
62 Å\textsuperscript{3}) was centred in a cubic box that was filled with
additional Adpn molecules: 600 (box 80 Å, 30,000 atoms) or 1200 (box 86
Å, 39,600 atoms), giving overall compositions of
(Adpn)\textsubscript{3}LiI and (Adpn)\textsubscript{4}LiI. Because of
periodicity, the grain is separated from its images by a layer of excess
Adpn on every face, a model of the Adpn-rich grain boundaries of the
non-stoichiometric samples. After energy minimization, the excess Adpn
was equilibrated around the frozen grain for 5 ns (NVT, 300 K), and the
whole system was then simulated without restraints for 10 ns (NpT, 300
K, 1 bar). The density reaches its plateau within 1 ns (boxes 73.7 and
79.9 Å), leaving boundary layers \textasciitilde1.1 and
\textasciitilde1.7 nm thick. Coordinates were saved every 1 ps.

\textbf{Table S8.} Diffusion coefficients (m\textsuperscript{2}
s\textsuperscript{-1}, medians over ions) in the grain-boundary models
at 300 K, in 1 ns windows before the grain core disorders. Grain
interior: caged ions more than 10 Å inside the grain. Boundary:
disordered ions. D\textsubscript{GB}/D\textsubscript{grain} is the ratio
of mean values; t\textsubscript{Li+} (NE) =
\ensuremath{\Sigma}D(Li\textsuperscript{+})/{[}\ensuremath{\Sigma}D(Li\textsuperscript{+}) +
\ensuremath{\Sigma}D(I\textsuperscript{-}){]} over the boundary ions.

\begin{longtable}[]{@{}
  >{\raggedright\arraybackslash}p{(\columnwidth - 16\tabcolsep) * \real{0.1656}}
  >{\raggedright\arraybackslash}p{(\columnwidth - 16\tabcolsep) * \real{0.0839}}
  >{\raggedright\arraybackslash}p{(\columnwidth - 16\tabcolsep) * \real{0.1104}}
  >{\raggedright\arraybackslash}p{(\columnwidth - 16\tabcolsep) * \real{0.1104}}
  >{\raggedright\arraybackslash}p{(\columnwidth - 16\tabcolsep) * \real{0.1104}}
  >{\raggedright\arraybackslash}p{(\columnwidth - 16\tabcolsep) * \real{0.1104}}
  >{\raggedright\arraybackslash}p{(\columnwidth - 16\tabcolsep) * \real{0.1104}}
  >{\raggedright\arraybackslash}p{(\columnwidth - 16\tabcolsep) * \real{0.0993}}
  >{\raggedright\arraybackslash}p{(\columnwidth - 16\tabcolsep) * \real{0.0993}}@{}}
\toprule\noalign{}
\begin{minipage}[b]{\linewidth}\raggedright
\textbf{Model}
\end{minipage} & \begin{minipage}[b]{\linewidth}\raggedright
\textbf{Window (ns)}
\end{minipage} & \begin{minipage}[b]{\linewidth}\raggedright
\textbf{Disordered Li\textsuperscript{+} / I\textsuperscript{-}}
\end{minipage} & \begin{minipage}[b]{\linewidth}\raggedright
\textbf{Grain interior D(Li\textsuperscript{+})}
\end{minipage} & \begin{minipage}[b]{\linewidth}\raggedright
\textbf{Grain interior D(I\textsuperscript{-})}
\end{minipage} & \begin{minipage}[b]{\linewidth}\raggedright
\textbf{Boundary D(Li\textsuperscript{+})}
\end{minipage} & \begin{minipage}[b]{\linewidth}\raggedright
\textbf{Boundary D(I\textsuperscript{-})}
\end{minipage} & \begin{minipage}[b]{\linewidth}\raggedright
\textbf{\shortstack{D(GB)/\\D(grain),\\mean}}
\end{minipage} & \begin{minipage}[b]{\linewidth}\raggedright
\textbf{\shortstack{t(Li\textsuperscript{+}),\\NE,\\boundary}}
\end{minipage} \\
\midrule\noalign{}
\endhead
\bottomrule\noalign{}
\endlastfoot
(Adpn)\textsubscript{3}LiI & 0--1 & 421 / 367 & 4.7 \ensuremath{\times}
10\textsuperscript{-13} & 4.2 \ensuremath{\times} 10\textsuperscript{-13} & 2.4 \ensuremath{\times}
10\textsuperscript{-11} & 2.2 \ensuremath{\times} 10\textsuperscript{-11} & 69 & 0.57 \\
(Adpn)\textsubscript{3}LiI & 1--2 & 507 / 484 & 3.9 \ensuremath{\times}
10\textsuperscript{-13} & 4.8 \ensuremath{\times} 10\textsuperscript{-13} & 3.1 \ensuremath{\times}
10\textsuperscript{-11} & 3.0 \ensuremath{\times} 10\textsuperscript{-11} & 46 & 0.54 \\
(Adpn)\textsubscript{4}LiI & 0--1 & 364 / 322 & 7.2 \ensuremath{\times}
10\textsuperscript{-13} & 6.8 \ensuremath{\times} 10\textsuperscript{-13} & 1.7 \ensuremath{\times}
10\textsuperscript{-11} & 2.1 \ensuremath{\times} 10\textsuperscript{-11} & 46 & 0.55 \\
(Adpn)\textsubscript{4}LiI & 1--2 & 497 / 471 & 6.8 \ensuremath{\times}
10\textsuperscript{-13} & 6.9 \ensuremath{\times} 10\textsuperscript{-13} & 2.6 \ensuremath{\times}
10\textsuperscript{-11} & 2.2 \ensuremath{\times} 10\textsuperscript{-11} & 43 & 0.55 \\
(Adpn)\textsubscript{4}LiI, annealed & 1--2 & 414 / 367 & 2.6 \ensuremath{\times}
10\textsuperscript{-13} & 2.3 \ensuremath{\times} 10\textsuperscript{-13} & 1.7 \ensuremath{\times}
10\textsuperscript{-11} & 1.3 \ensuremath{\times} 10\textsuperscript{-11} & 86 & 0.55 \\
(Adpn)\textsubscript{4}LiI, annealed & 2--3 & 472 / 436 & 3.0 \ensuremath{\times}
10\textsuperscript{-13} & 4.7 \ensuremath{\times} 10\textsuperscript{-13} & 2.3 \ensuremath{\times}
10\textsuperscript{-11} & 2.4 \ensuremath{\times} 10\textsuperscript{-11} & 46 & 0.53 \\
\end{longtable}

\subsection{Diffusion in the grain interior and in the
boundaries}\label{diffusion-in-the-grain-interior-and-in-the-boundaries}

While the grain core is intact, the per-ion diffusion coefficients split
into two populations \textbf{(Figure S17)}. Caged ions show the cage
vibrations of the bulk crystal at 300 K (Figure S10a; grain-interior
medians 2--7 \ensuremath{\times} 10\textsuperscript{-13} m\textsuperscript{2}
s\textsuperscript{-1}), while disordered ions near the boundary diffuse
at 1.3--3.1 \ensuremath{\times} 10\textsuperscript{-11} m\textsuperscript{2}
s\textsuperscript{-1} (medians), close to the \textsuperscript{7}Li
PFG-NMR values (dotted lines). The ratio of the mean values is 43--86
(Table S8). With 5--10\% of the ions in the boundary regions, as
estimated from the static \textsuperscript{7}Li line shape, the
boundaries carry 69--91\% of the Nernst--Einstein conductance (Figure
S18e).

\includegraphics[width=\textwidth,height=0.82\textheight,keepaspectratio]{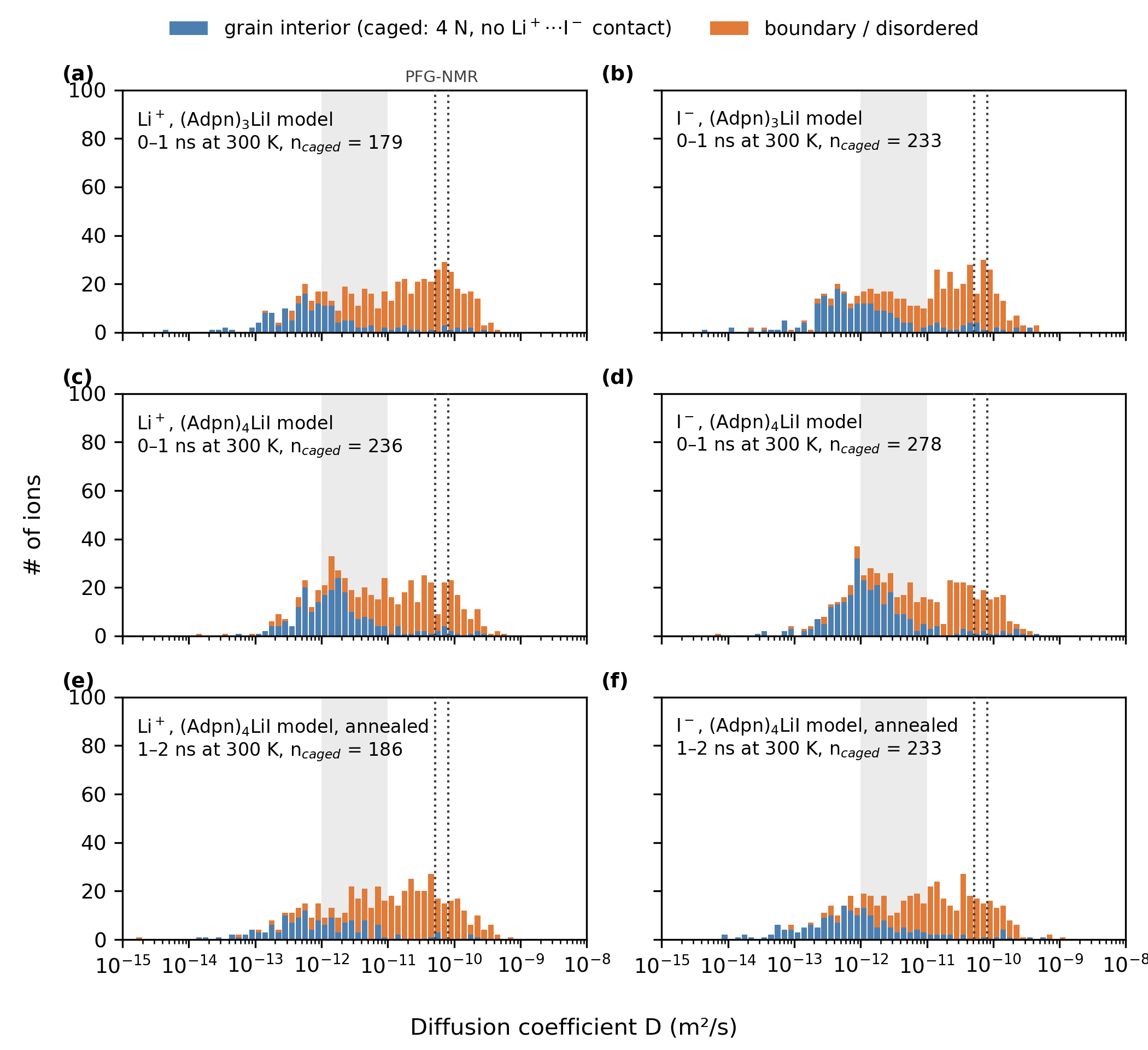}

\textbf{Figure S17.} Distribution of the diffusion coefficients of
Li\textsuperscript{+} (a, c, e) and I\textsuperscript{-} (b, d, f) ions
in the grain-boundary models at 300 K: (a, b)
(Adpn)\textsubscript{3}LiI, 0--1 ns; (c, d) (Adpn)\textsubscript{4}LiI,
0--1 ns; (e, f) annealed (Adpn)\textsubscript{4}LiI, 1--2 ns. Bars are
stacked: grain-interior (caged) ions in blue, boundary (disordered) ions
in orange. Gray area as in Figures S10 and S11; dotted lines,
\textsuperscript{7}Li PFG-NMR diffusion coefficients (5.15 \ensuremath{\times}
10\textsuperscript{-11} and 8.1 \ensuremath{\times} 10\textsuperscript{-11}
m\textsuperscript{2} s\textsuperscript{-1}).

\includegraphics[width=\textwidth,height=0.82\textheight,keepaspectratio]{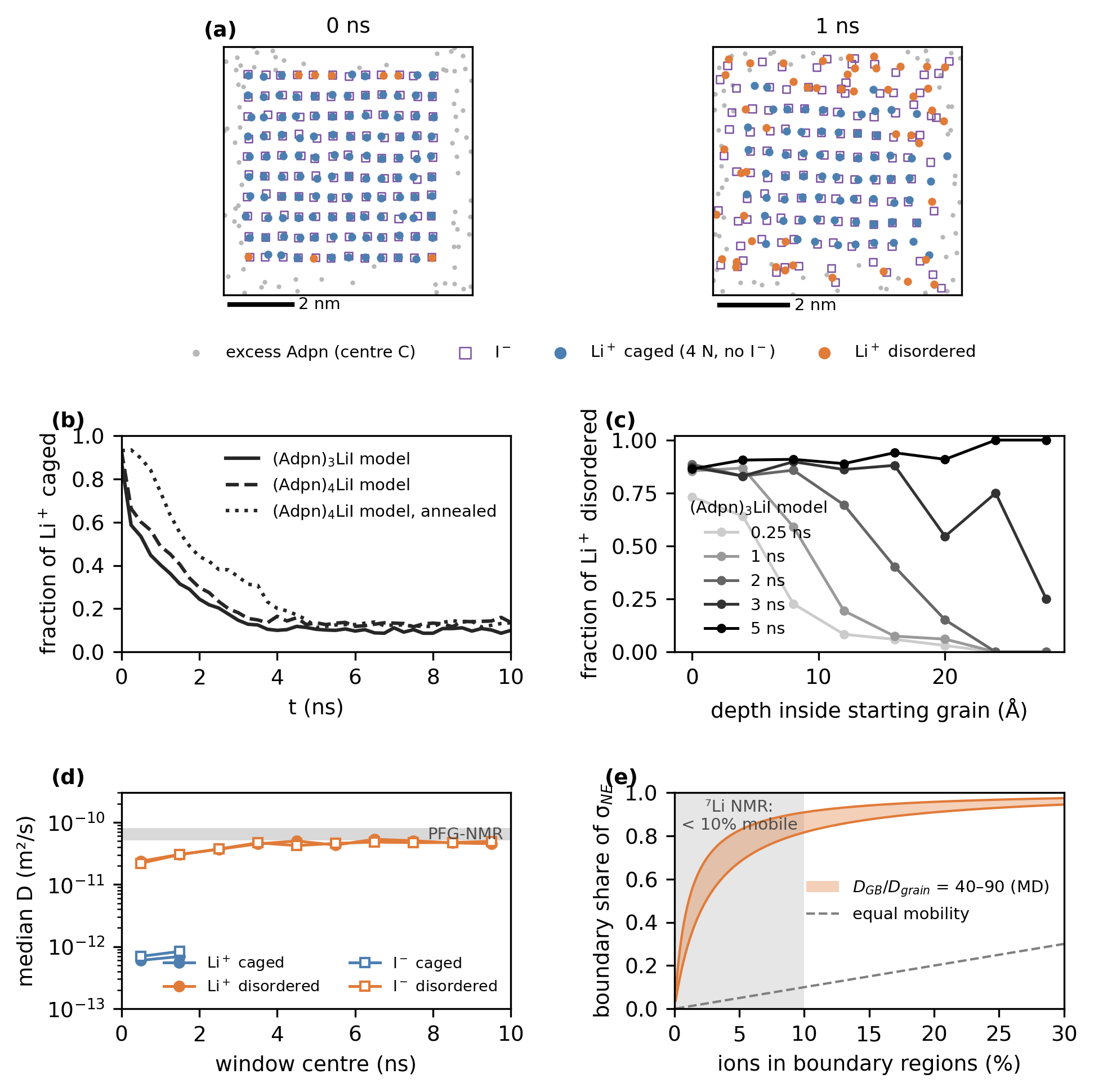}

\textbf{Figure S18.} Grain-boundary model dynamics at 300 K. (a) 12 Å
slice through the centre of the (Adpn)\textsubscript{3}LiI model at 0
and 1 ns: caged Li\textsuperscript{+} (blue), disordered
Li\textsuperscript{+} (orange), I\textsuperscript{-} (open squares) and
the central carbon of the excess Adpn (gray). (b) Fraction of caged
Li\textsuperscript{+} versus time in the three runs. (c) Fraction of
disordered Li\textsuperscript{+} versus depth inside the starting grain,
(Adpn)\textsubscript{3}LiI model. (d) Median diffusion coefficients of
caged (blue) and disordered (orange) Li\textsuperscript{+} (filled) and
I\textsuperscript{-} (open) in successive 1 ns windows,
(Adpn)\textsubscript{3}LiI model; caged values are shown while at least
50 ions remain caged; gray band, \textsuperscript{7}Li PFG-NMR. (e)
Boundary share of the Nernst--Einstein conductance versus the fraction
of ions in the boundary regions, for the measured mobility ratio
D\textsubscript{GB}/D\textsubscript{grain} = 40--90; gray area, boundary
fraction below 10\% from \textsuperscript{7}Li NMR; dashed line, equal
mobilities.

\subsection{Limitations}\label{limitations-1}

\begin{enumerate}
\def\labelenumi{(\roman{enumi})}
\item
  Over 1 ns the caged ions only vibrate, so their D values are upper
  bounds and the ratios D\textsubscript{GB}/D\textsubscript{grain} are
  lower bounds.
\item
  The grain (\textasciitilde6 nm) is much smaller than the crystallites
  of the samples, so the model boundary fraction (\textgreater25\%) is
  not that of the samples; the extrapolation in Figure S18e uses the
  boundary fraction from \textsuperscript{7}Li NMR and assumes
  continuous boundary paths.
\item
  Only the LT grain at 300 K was simulated, an HT grain was not.
\end{enumerate}

\section*{References}

(1) Bruce, P. G.; Vincent, C. A. Polymer Electrolytes. \emph{J. Chem.
Soc. Faraday Trans.} \textbf{1993}, \emph{89} (17), 3187.
https://doi.org/10.1039/ft9938903187

(2) Stejskal, E. O.; Tanner, J. E. Spin Diffusion Measurements: Spin
Echoes in the Presence of a Time-Dependent Field Gradient. \emph{J.
Chem. Phys.} \textbf{1965}, \emph{42} (1), 288--292.
https://doi.org/10.1063/1.1695690

(3) Giannozzi, P.; Andreussi, O.; Brumme, T.; Bunau, O.; Buongiorno
Nardelli, M.; Calandra, M.; Car, R.; Cavazzoni, C.; Ceresoli, D.;
Cococcioni, M.; Colonna, N.; Carnimeo, I.; Dal Corso, A.; de Gironcoli,
S.; Delugas, P.; DiStasio, R. A., Jr.; Ferretti, A.; Floris, A.;
Fratesi, G.; Fugallo, G.; Gebauer, R.; Gerstmann, U.; Giustino, F.;
Gorni, T.; Jia, J.; Kawamura, M.; Ko, H. Y.; Kokalj, A.; Kucukbenli, E.;
Lazzeri, M.; Marsili, M.; Marzari, N.; Mauri, F.; Nguyen, N. L.; Nguyen,
H. V.; Otero-de-la-Roza, A.; Paulatto, L.; Ponce, S.; Rocca, D.;
Sabatini, R.; Santra, B.; Schlipf, M.; Seitsonen, A. P.; Smogunov, A.;
Timrov, I.; Thonhauser, T.; Umari, P.; Vast, N.; Wu, X.; Baroni, S.
Advanced Capabilities for Materials Modelling with Quantum ESPRESSO.
\emph{J. Phys. Condens. Matter} \textbf{2017}, \emph{29} (46), 465901.
https://doi.org/10.1088/1361-648X/aa8f79

(4) Breneman, C. M.; Wiberg, K. B. Determining Atom-Centered Monopoles
from Molecular Electrostatic Potentials. The Need for High Sampling
Density in Formamide Conformational Analysis. \emph{J. Comput. Chem.}
\textbf{1990}, \emph{11} (3), 361--373.
https://doi.org/10.1002/jcc.540110311

(5) Neese, F. Software Update: The ORCA Program System---Version 6.0.
\emph{WIREs Comput. Mol. Sci.} \textbf{2025}, \emph{15} (2), e70019.
https://doi.org/10.1002/wcms.70019

(6) Prakash, P.; Fall, B.; Aguirre, J.; Sonnenberg, L. A.; Chinnam, P.
R.; Chereddy, S.; Dikin, D. A.; Venkatnathan, A.; Wunder, S. L.; Zdilla,
M. J. A Soft Co-Crystalline Solid Electrolyte for Lithium-Ion Batteries.
\emph{Nat. Mater.} \textbf{2023}, 1--9.
https://doi.org/10.1038/s41563-023-01508-1

(7) Abraham, M. J.; Murtola, T.; Schulz, R.; Páll, S.; Smith, J. C.;
Hess, B.; Lindahl, E. GROMACS: High Performance Molecular Simulations
through Multi-Level Parallelism from Laptops to Supercomputers.
\emph{SoftwareX} \textbf{2015}, \emph{1--2}, 19--25.
https://doi.org/10.1016/j.softx.2015.06.001

(8) Sheppard, D.; Xiao, P.; Chemelewski, W.; Johnson, D. D.; Henkelman,
G. A Generalized Solid-State Nudged Elastic Band Method. \emph{J. Chem.
Phys.} \textbf{2012}, \emph{136}, 074103.
https://doi.org/10.1063/1.3684549

(9) Larsen, A. H.; Mortensen, J. J.; Blomqvist, J.; Castelli, I. E.;
Christensen, R.; Dułak, M.; Friis, J.; Groves, M. N.; Hammer, B.;
Hargus, C.; Hermes, E. D.; Jennings, P. C.; Jensen, P. B.; Kermode, J.;
Kitchin, J. R.; Kolsbjerg, E. L.; Kubal, J.; Kaasbjerg, K.; Lysgaard,
S.; Maronsson, J. B.; Maxson, T.; Olsen, T.; Pastewka, L.; Peterson, A.;
Rostgaard, C.; Schiøtz, J.; Schütt, O.; Strange, M.; Thygesen, K. S.;
Vegge, T.; Vilhelmsen, L.; Walter, M.; Zeng, Z.; Jacobsen, K. W. The
Atomic Simulation Environment---a Python Library for Working with Atoms.
\emph{J. Phys. Condens. Matter} \textbf{2017}, \emph{29} (27), 273002.
https://doi.org/10.1088/1361-648X/aa680e

(10) Batatia, I.; Kovács, D. P.; Simm, G. N. C.; Ortner, C.; Csányi, G.
MACE: Higher Order Equivariant Message Passing Neural Networks for Fast
and Accurate Force Fields. \emph{Adv. Neural Inf. Process. Syst.}
\textbf{2022}, \emph{35}, 11423--11436.

(11) Batatia, I.; Benner, P.; Chiang, Y.; Elena, A. M.; Kovács, D. P.;
Riebesell, J.; Advincula, X. R.; Asta, M.; Avaylon, M.; Baldwin, W. J.;
Berger, F.; Bernstein, N.; Bhowmik, A.; Bigi, F.; Blau, S. M.; Cărare,
V.; Ceriotti, M.; Chong, S.; Darby, J. P.; De, S.; Della Pia, F.;
Deringer, V. L.; Elijošius, R.; El-Machachi, Z.; Falcioni, F.; Fako, E.;
Ferrari, A. C.; Gardner, J. L. A.; Gawkowski, M. J.; Genreith-Schriever,
A.; George, J.; Goodall, R. E. A.; Grandel, J.; Grey, C. P.; Grigorev,
P.; Han, S.; Handley, W.; Heenen, H. H.; Hermansson, K.; Holm, C.; Ho,
C. H.; Hofmann, S.; Jaafar, J.; Jakob, K. S.; Jung, H.; Kapil, V.;
Kaplan, A. D.; Karimitari, N.; Kermode, J. R.; Kourtis, P.; Kroupa, N.;
Kullgren, J.; Kuner, M. C.; Kuryla, D.; Liepuoniute, G.; Lin, C.;
Margraf, J. T.; Magdău, I.-B.; Michaelides, A.; Moore, J. H.; Naik, A.
A.; Niblett, S. P.; Norwood, S. W.; O'Neill, N.; Ortner, C.; Persson, K.
A.; Reuter, K.; Rosen, A. S.; Rosset, L. A. M.; Schaaf, L. L.; Schran,
C.; Shi, B. X.; Sivonxay, E.; Stenczel, T. K.; Svahn, V.; Sutton, C.;
Swinburne, T. D.; Tilly, J.; van der Oord, C.; Vargas, S.;
Varga-Umbrich, E.; Vegge, T.; Vondrák, M.; Wang, Y.; Witt, W. C.; Wolf,
T.; Zills, F.; Csányi, G. A Foundation Model for Atomistic Materials
Chemistry. \emph{arXiv} \textbf{2023}, 2401.00096.

(12) Perdew, J. P.; Burke, K.; Ernzerhof, M. Generalized Gradient
Approximation Made Simple. \emph{Phys. Rev. Lett.} \textbf{1996},
\emph{77}, 3865--3868. https://doi.org/10.1103/PhysRevLett.77.3865

(13) Grimme, S.; Ehrlich, S.; Goerigk, L. Effect of the Damping Function
in Dispersion Corrected Density Functional Theory. \emph{J. Comput.
Chem.} \textbf{2011}, \emph{32}, 1456--1465.
https://doi.org/10.1002/jcc.21759

(14) Kresse, G.; Furthmüller, J. Efficient Iterative Schemes for Ab
Initio Total-Energy Calculations Using a Plane-Wave Basis Set.
\emph{Phys. Rev. B} \textbf{1996}, \emph{54}, 11169--11186.
https://doi.org/10.1103/PhysRevB.54.11169

(15) Smidstrup, S.; Pedersen, A.; Stokbro, K.; Jónsson, H. Improved
Initial Guess for Minimum Energy Path Calculations. \emph{J. Chem.
Phys.} \textbf{2014}, \emph{140}, 214106.
https://doi.org/10.1063/1.4878664

(16) Bitzek, E.; Koskinen, P.; Gähler, F.; Moseler, M.; Gumbsch, P.
Structural Relaxation Made Simple. \emph{Phys. Rev. Lett.}
\textbf{2006}, \emph{97}, 170201.
https://doi.org/10.1103/PhysRevLett.97.170201

(17) Henkelman, G.; Uberuaga, B. P.; Jónsson, H. A Climbing Image Nudged
Elastic Band Method for Finding Saddle Points and Minimum Energy Paths.
\emph{J. Chem. Phys.} \textbf{2000}, \emph{113}, 9901--9904.
https://doi.org/10.1063/1.1329672

(18) Prakash, P. Drop, Render, Spin, Trace, Analyze - DRSTA 0.1.0, 2026.
https://ppt-2.github.io/drsta/

\end{document}